\documentstyle[aps,twocolumn]{revtex}
\input psfig

\begin{document}

\begin{center}
{\bf \Large Cranked shell model and isospin symmetry near $N=Z$\\ }
\bigskip
{\bf \large Stefan G. Frauendorf$^{(1)}$ and Javid A. Sheikh$^{(1,2,3)}$}\\
\bigskip
{\it $^{(1)}$IKHP, Research Center Rossendorf, PF 510119, 
D-01314 Dresden, Germany\\
$^{(2)}$Physik-Department, Technische Universit\"at M\"unchen,
85747, Garching, Germany\\
$^{(3)}$Tata Institute of Fundamental Research, Bombay,
400 005, India}
\end{center}


\begin{abstract}
A cranked shell model  approach for the description of rotational bands in 
$N\approx Z$ nuclei is formulated. The isovector
neutron-proton
pairing is taken into account explicitly.  
The concept of spontaneous breaking and subsequent
restoration of the isospin symmetry turns out to be crucial.
The general rules to construct the near yrast-spectra for rotating nuclei are
presented. 
For the model case of particles in a j-shell,
it is shown that excitation spectra and 
the alignment processes are well  described as compared to the
exact shell model calculation. Realistic cranked shell model
calculations are able to 
describe the experimental spectra of $^{72,73}$Kr and $^{74}$Rb isotopes.
\end{abstract}

{\it keywords: proton neutron pairing, rotational spectra, isospin symmetry}
\section{Introduction}

The classification of rotational bands as quasiparticle configurations in
a rotating mean-field  has led to an  understanding of the  yrast-
spectra of rapidly rotating nuclei \cite{ring,bf79}. This popular approach
referred to as the cranked shell model (CSM) has been employed quite
extensively and is reviewed in refs. 
\cite{devoigt,bg}. 
These high-spin studies have provided new insights into the nature of the 
pair-correlations among identical particles \cite{pairing}.  
The CSM in its traditional form assumes that there are no 
proton-neutron (pn-) pair correlations. 
Modern $\gamma$-ray detector arrays
will allow to study high-spin states for nuclei near the $N=Z$ line in the 
mass $\sim$ 50 and 80 regions \cite{len96,lea97,kr72} and these possibilities 
will be greatly enhanced with the
availability of the radioactive beams. 
For these nuclei strong pn-pair correlations
are expected. By studying the rotational bands one may
obtain new information about the pn-pairing in deformed nuclei. 
For this purpose, it is
necessary to reformulate the CSM in such a way that the pn-pairing is
included.

The main motivation of the present work is to develop a quasiparticle
CSM approach near $N=Z$ which includes the pn-pairing effects.
As will be shown, the isospin symmetry plays a central role,
permitting to derive the basic structure of the rotational spectra
in terms of quasiparticle configurations
in a $T=1$ pair-field. General rules for
constructing the quasiparticle excitation spectra in the presence of np-pairing
will be provided.
We shall study a model problem of protons and neutrons in a 
deformed $f_{7/2}$ shell interacting with a $\delta$-force, 
which is solved by exact numerical diagonalization.
The suggested CSM approach will be tested using these exact results. We have
used the $f_{7/2}$ shell for the simplicity in carrying out the exact deformed
shell model (SM) calculations, expecting the CSM predictions to improve in 
larger shells as the mean-field becomes more dominant
with increasing number of particles.

It will be demonstrated that the suggested 
CSM with good isospin can be cast into
a form that permits calculations 
by means of a conventional CSM approach.
Taking advantage of this, the yrast-spectra of even-even,
odd-odd $N=Z$ and odd-$A$ $N=Z\pm 1$ nuclei near $A=72$ are constructed and  
compared with the  experimental data. The meager experimental
information available is consistent with the suggested dominance of a $T=1$
monopole pn-pair field.

\section{Deformed shell model}

As a case study for the CSM to be developed, we use
the deformed shell model hamiltonian which 
consists of a cranked deformed
one-body term, $h^\prime$ and a scalar two-body delta-interaction 
\cite{srn90,she90,fsr94}. The one-body term is the familiar
cranked Nilsson mean-field potential which takes into account of the
long range part of the nucleon-nucleon interaction. The residual short range
interaction is specified by the delta-interaction. In many of the high spin
studies, the residual interaction is in the form of a monopole interaction.
However, it has been demonstrated \cite{mueller} that the higher multipoles 
can be important. Hence, in our model study
we consider all possible multipole components of the delta-interaction
and also all the possible interaction terms, proton-proton (pp), 
neutron-neutron (nn)
and  neutron-proton (np).
The SM hamiltonian employed  is given by
\begin{equation} \label{H}
H^\prime = h^\prime -g \delta(\hat r_1 - \hat r_2)
\end{equation}
where,
\begin{equation}\label{h'}
h^\prime = -4 \kappa { \sqrt { 4 \pi \over 5 }} Y_{20}  - \omega J_x.
\end{equation}
We use 
$G=g\int R^4_{nl}r^2 dr$ as our energy unit and the deformation
energy $\kappa$ is related to the deformation parameter $\beta$.
For the case of $f_{7/2}$ shell, $\kappa$=1.75 approximately
corresponds to $\beta=0.16$. In order to solve the eigenvalue problem exactly,
we are limited to a small configuration space. As in the previous work,
the model space in the present analysis consists of a single j shell.
We have diagonalized the hamiltonian (\ref{H}) exactly for neutrons and protons
in the $f_{7/2}$ shell. As the strengths of nn-, pp- and np-parts 
are identical, the hamiltonian  is
invariant with respect to rotations in isospace, i. e.
\begin{equation}
{\cal R}H^\prime {\cal R}^{-1}=H^\prime,
\end{equation}  
where ${\cal R}$ defines a rotation in isospace, generated by the 
isospin operators $T_x,~T_y$ and $T_z$.
Furthermore, the hamiltonian (1) is invariant with respect to a 
spatial rotation
about the x-axis by an angle of $\pi$. As a consequence, the signature $\alpha$
is a good quantum number \cite{bf79}, which implies that the SM solutions
represent states with the angular-momentum $I=\alpha +2n$ ($n$ integer).

\section{Mean-field approach}
As mentioned in the introduction, the motivation of the present work is
to develop a mean-field approach in the presence of the pn-pairing. The
mean-field CSM approach in the case of identical particles 
has been quite successful to describe the rotational properties of medium
and heavy mass nuclei. In the simplest form of this approach, the mean-fields
in the form of the  pair-gap and the deformation are held fixed as a 
function of the rotational frequency. In a fully selfconsistent 
 Hartree-Fock-Bogolubov (HFB) calculation, the mean-field 
(most important the pair-field )  often
shows rapid changes  as a function of rotational frequency that are 
due to the  broken symmetries in the intrinsic
frame of reference. The rapid changes  are smeared out by projecting
out the wavefunction with good quantum numbers (most important 
the particle number). In the CSM approach  the problem of 
the ``phase transition'' is avoided, because  the mean-fields are held
fixed. This leads, often, to a better description than  
a selfconsistent HFB calculation \cite{pnr89}. However, we would like 
to caution here that this approach has obvious limitations.
It holds below a phase transition and permits a fair description
of the lower part of the extended transition region.
With the appropriate choice of the mean-field, the same is true for 
the upper region. Thus used with care,  the CSM is a robust and a simple 
tool to analyze the near-yrast spectra.

In subsection \ref{hfb.sec}, we briefly present the HFB approach
in the presence of the np-pairing. There is an extensive 
literature on this subject that has been reviewed in ref.
\cite{goodman}, to which we refer concerning the previous work.  
The HFB equations are solved
selfconsistently in subsection \ref{hfbj.sec}. It will be found that for low-
frequency, $\omega$ the pair-field has isovector character. Using this
 result and other studies \cite{sandhu,satula,engel,dean}, which also
suggest the dominance of isovector pairing at low-$\omega$,  we proceed
with the assumption that for realistic nuclei above
mass=40, the pair-field has isovector character. The CSM for such
a scenario is developed in subsections
\ref{sbiso.sec}-\ref{csm.sec}, where 
the general rules for constructing
the quasiparticle spectra are given.

\subsection{Hartree-Fock-Bogolubov equations}\label{hfb.sec}

In the development of the CSM with np-pairing, we have
employed the HFB method,  details of which can be
found, for instance, in refs.\cite{goodman,ringschuck}.
The HFB equations are given by
\begin{equation}\label{hfb}
{\cal H}'\left( \begin{array}{c} U\\V \end{array}\right)=e'_i \left( 
\begin{array}{c} U\\V \end{array}\right)
\end{equation}

where
\begin{eqnarray}
{\cal H}'=&
\left( \begin{array}{ccc} h'_{ij}+\Gamma_{ij}&
-(\lambda+\lambda_\tau \tau_i)\delta_{ij}& \Delta_{ij}\\
-\Delta_{ij}^{*}&-h'_{ij}-\Gamma_{ij}&
+(\lambda+\lambda_\tau \tau_i)\delta_{ij} 
\end{array}\right)\label{hfbh}&\\
\Gamma_{ij} =& \sum_{kl} \langle ik|v_a|jl\rangle  \rho_{lk},& \label{hfbg}\\
\Delta_{ij} =& {1 \over 2} \sum_{kl} \langle  ij|v_a|kl\rangle  t_{kl},&
 \label{hfbd}\\
\rho   =& V^{*} V^{T},&\label{hfbr}\\
 t     =& V^{*} U^{T}. &\label{hfbt}
\end{eqnarray}
The quantities in the brackets $\langle v_a\rangle $ in (\ref{hfbg}) 
and (\ref{hfbd})
are the antisymmetric uncoupled
matrix elements of the interaction.
In Eq. (\ref{hfbh}), we have introduced the isospin label $\tau =1, -1$ for neutrons
and protons, respectively and rearranged the chemical potentials 
$\lambda_n$ and $\lambda_p$, which constrain $N$ and $Z$, into 
$\lambda=(\lambda_n+ \lambda_p)/2$ and 
$\lambda_\tau=\lambda_n- \lambda_p$ which fix mass $A$ and the 
isospin projection $T_z$, respectively.  
The HFB solutions are obtained by solving the equations (\ref{hfb})
-(\ref{hfbt})
selfconsistently.

The pairs of states  $\{ ij\}$ that define the pair-field (\ref{hfbd})
can be rewritten in a coupled representation  as $\{ t,t_z,\alpha,\beta\}$, 
which explicitly indicates the isospin $t$ and $t_z$ and  $\alpha,\beta$
denote all other quantum numbers except the isospin.  For the single-$j$ shell
model, $\alpha,\beta=J,M$, where
$J,M$ are the angular-momentum and its projection, 
which are sufficient 
to fix the  quantum numbers. 
If $t=0$, the pair-field is an isoscalar and for $(t=1,t_z)$ it is
an isovector.
The pp-pair-field has $(t=1, t_z=-1)$ and the nn-
has $(t=1, t_z=1)$. There are two pn-pair-fields with
$(t=1, t_z=0)$ and $(t=0, t_z=0)$. We use the lower case letters 
$t$ and $t_z$ for the isospin
of the pair-field in order to 
avoid confusion  with the total isospin of the states, which we denote
by $T$ and $T_z$. 

It is known that  in order to treat the $t=0$ and the $t=1$ 
fields simultaneously, it is 
necessary to define complex HFB potentials \cite{cg,T0}. The 
$t=1$ and $t=0$
pair-fields correspond to the real and imaginary parts of the complex 
pair-potential $\Delta$. 
The initial complex HFB wave functions are constructed by using
the expressions for real and imaginary $V$'s and $U$'s of the HFB 
transformation
in terms of the pair-gaps given in ref. \cite{goodgosw}. 
No symmetry has been imposed on the
HFB wavefunction since it is known that this may lead to exclusion
of particular correlations \cite{mueller}. 

\subsection{HFB solution for the $j$-shell}\label{hfbj.sec}

We find that the pair-field
is {\em either} $t=1$ {\em or} $t=0$. This coincides with earlier results
of refs. \cite{mueller,satula}, where in contrast to the present work, 
 additional symmetries were imposed. Hence, the mutual exclusiveness of the 
$t=1$ and $t=0$ pair-fields is not a consequence of these
symmetries. The lack of a mixed phase can be interpreted qualitatively
in the following way:
Since the interaction conserves isospin, it scatters pairs only into
pairs with the same $t$. The correlation energy is of the order of
$ n^2$, with $n$
being the number of states the pairs can be scattered into.
Since the $t=1$ pairs block partially the $t=0$ phase space, and vice versa,
the pure field has a larger correlation energy.

For the $(N=Z=4)$ system we find a $t=1$ pair-field at low-rotational
frequency $\omega$. Fig. \ref{jx.fig} shows that there is a rapid alignment
at $\omega = 0.45G$, where the self consistent solution changes to a $t=0$
pair-field.
For the $(N=Z=3)$ system we also find  a $t=1$ pair-field at low-
$\omega$, which changes to a $t=0$ field at a higher frequency. 
The change is
associated with the crossing of the $T=1$ g-band with an aligned $T=0$
odd-spin band, which will be discussed below. The change from 
a $t=1$ to a $t=0$ field has also been found in the HFB calculation of ref. 
\cite{satula} when changing the $t=0$ strength of the np-interaction.  
In the case of the $t=1$ solution, the $J\not=0$ fields are found to be
small as compared to
the monopole pair-field, $J=0$.
In the remainder of the paper, we shall  
consider the various aspects of the $t=1$ HFB solutions.
The $t=0$ solutions have been studied separately
\cite{sw,fst0}. We only mention that the $J=1$ and $7$ components of the  
pair-fields dominate \cite{sw}. 

As compared to the SM calculation, the fully selfconsistent HFB solution
shows a much too early alignment and has a wrong behavior after the
crossing. This is related to the fact that $t=1$ pair-field vanishes
at the crossing point. The crossings frequency
 in even-even nuclei is determined by
competition between the $t=1$ pair-field and the Coriolis force. In
the exact calculations, the $t=1$ pair-correlations persist to very high
frequencies \cite{sw}.

\subsection{Spontaneous breaking of the isospin symmetry by 
the $t=1$ pair-field}\label{sbiso.sec}
 
Before discussing the symmetry breaking by the isovector pair-field, it is
useful to state the familiar case of spontaneous breaking of the 
spatial isotropy by a  mean-field solution with a deformed density
distribution (c.f. ref.\cite{ringschuck,bm2}). Since the two-body
hamiltonian is isotropic, this symmetry is broken spontaneously. 
There is a family of mean-field solutions with the same energy which 
correspond to different orientations of the density distribution.
All represent one and the same intrinsic quasiparticle configuration,
which is not
an eigenfunction of the total angular-momentum. Any of these solutions can be 
chosen as the intrinsic state. The principal axes of its density
distribution define the body-fixed coordinate system.
The states of good angular momentum are superpositions of these states of
different orientations, the weight being given by the Wigner
$D$-functions. Thus, the relative importance of the different orientations
is fixed by restoring the angular momentum symmetry. At the simplest level
of the cranking model, which is valid for sufficiently strong symmetry 
breaking, 
the energy of the good angular momentum
state is given by the mean-field value.      

\begin{figure}[t]
\mbox{\psfig{file=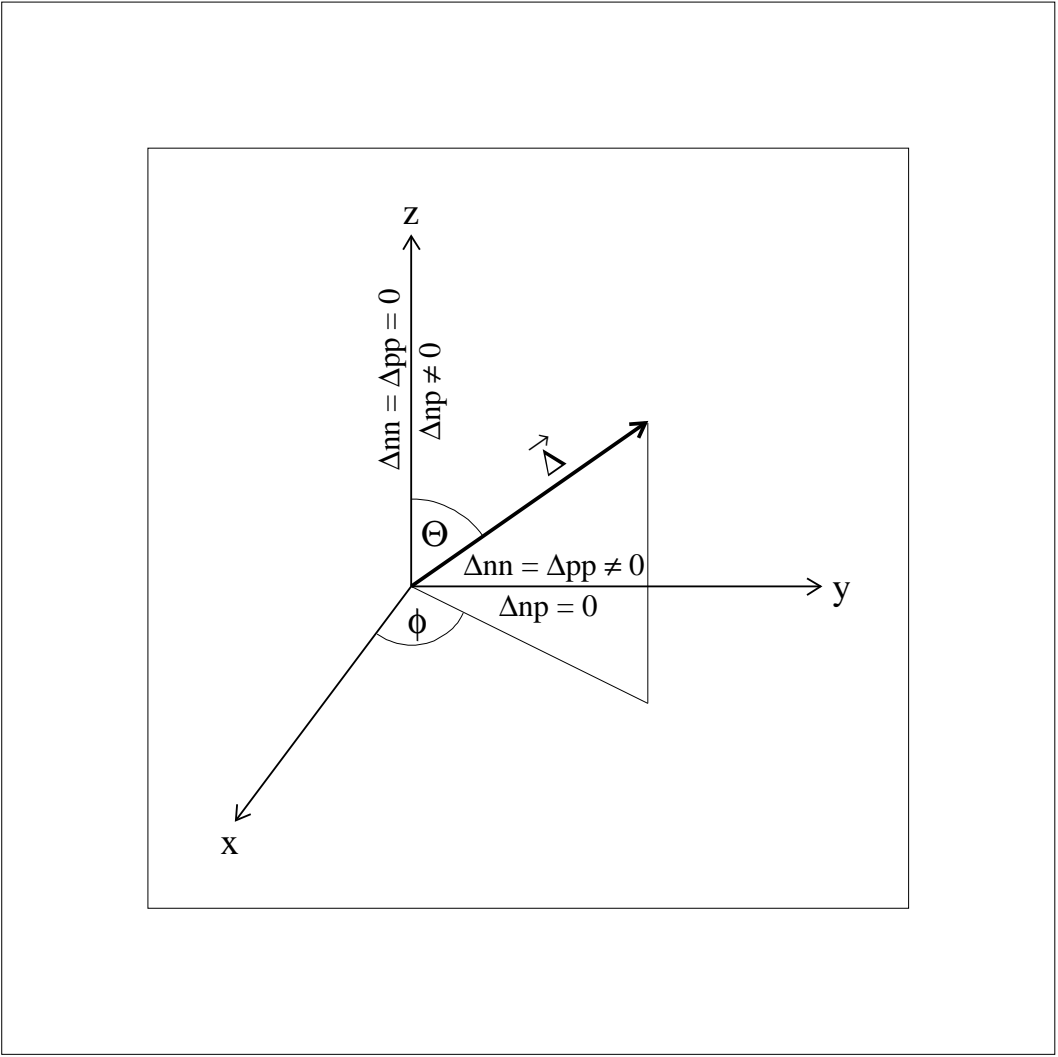,width=14cm}}
\caption{\label{delta.fig}
The isovector pair-field $\vec \Delta$.}
\end{figure}

Let us now consider a $t=1$ HFB solution found for the $N=Z$ system (
for example, the one discussed  in the
previous subsection).  
The $t=1$ pair-field $\vec \Delta$ is a vector that
points in a certain direction in isospace,  breaking the 
isospin symmetry. Since the two body hamiltonian is isospin invariant, 
 the symmetry is a spontaneously broken and all orientations of the 
isovector pair-field : 
\begin{eqnarray}
\Delta_{J,M,t=1,t_z=\pm 1}=\mp \Delta_{J,M,t=1}
 \sin{\theta}\exp{\mp i \phi}/\sqrt{2} \nonumber \\
\Delta_{J,M,t=1,t_z=0}=\Delta_{J,M,t=1} \cos{\theta}
\end{eqnarray}
are equivalent. Fig. \ref{delta.fig} illustrates this family of  HFB 
 solutions, the energy of which 
does not depend on the orientation of the pair-field.
In particular, the cases of a pure pn-field ($\theta = 0$, z-axis) and 
pure pp- and nn-pair fields ($\theta= \pi/2$, $\phi=0$, y-axis) represent
the same intrinsic state. Hence, on the mean-field level
the ratio between the strengths of  pp-, nn- and pn-pair-fields is given by
the orientation of the pair-field, which is not determined by the HFB
procedure.
The relative strengths 
of the three types of pair-correlations becomes
only definite when the isospin symmetry is restored.
The symmetry breaking by the isovector pair-field has been discussed 
before in  \cite{camiz,ginocchio}, where references to
earlier work can be found. 

\subsection{Intrinsic excitations with $T=0$}\label{t=0.sec}

Let us first discuss the 
states with total isospin $T=0$. They are isotropic 
superpositions of all the  orientations
of the pair-field, which  corresponds to an equal amount of pn-, pp- and 
nn-correlation energy. 

Like in the case of spatial rotation, the intrinsic excitations are constructed
from the quasiparticles (qps) belonging to {\em one} of the orientations of pair-field.
We choose the y-direction, $\Delta_{nn}=\Delta_{pp}, \Delta_{np}=0$. 
This is a particularly  convenient choice because it permits to
reduce the construction
of the qp-excitation spectrum to the familiar case with no pn-pairing 
\cite{bf79}. The choice of the qp operators is not unique \cite{ginocchio}. 
We choose them to be pure quasineutrons or quasiprotons and 
denote their creation operators by $\beta_{t_z,k}^+$,
where $t_z$  indicates  the isospin projection. They are pairwise degenerate,
i.e. the qp routhians $e'(\omega)_{\frac{1}{2},k}=e'(\omega)_{-\frac{1}{2},k}$
are equal\footnote{This degeneracy permits to use linear combinations 
of the two
degenerate qp as new qp operators, which, thus, are not uniquely
determined. Another choice will be discussed below.}. This choice has the advantage that
\begin{equation}
[ e^{-i\pi Z},{\cal H}'] = [ e^{-i\pi N},{\cal H}'] =0,
\end{equation}
and particle number parity becomes a good quantum number.
The HFB vacuum state has $N$ and $Z$ even. 
Configurations with an odd or even-number of quasineutrons belong to the 
odd- or even-$N$, respectively, and a similar thing holds for the protons.

However, not all qp configurations are permitted. If $\lambda_\tau=0$,
the qp routhian   commutes with $T_y$, 
\begin{equation}
[ T_y,{\cal H}'] = 0.
\end{equation}
This implies that the qp configurations
have  $T_y$ as a good quantum number. Since $T \geq T_y$, only
 configurations with $T_y=0$ are permitted. The construction of these
configurations may 
be based on the following relations for the qp operators
\begin{equation}\label{com1qp}
[T_+,\beta_{\frac{1}{2},k}^+]=0,~~~ [T_+,\beta_{-\frac{1}{2},k}^+]=
\beta_{\frac{1}{2},k}^+, 
\end{equation}
\begin{equation}
[T_-,\beta_{\frac{1}{2},k}^+]=\beta_{-\frac{1}{2},k}^+,~~~
 [T_-,\beta_{-\frac{1}{2},k}^+]=0. 
\end{equation}
The qp vacuum has $T_y=0$. This can been seen using its representation in the 
form \cite{ringschuck}
\begin{equation}
|0\rangle\propto \prod \beta_{\frac{1}{2},k}\beta_{-\frac{1}{2},k}|00\rangle.
\end{equation}
The relations (\ref{com1qp}) imply
\begin{equation}\label{com11qp}
[T_+,\beta_{\frac{1}{2},k}^+\beta_{-\frac{1}{2},k}^+]=
[T_-,\beta_{\frac{1}{2},k}^+\beta_{-\frac{1}{2},k}^+]=
[T_y,\beta_{\frac{1}{2},k}^+\beta_{-\frac{1}{2},k}^+]=0
\end{equation}
and the same for the annihilation operators. From (\ref{com1qp}) and
(\ref{com11qp}) it follows that 
\begin{equation}
T_y|0\rangle =0, 
~~~ T_y\beta_{\frac{1}{2},k}^+\beta_{-\frac{1}{2},k}^+|0\rangle =0. 
\end{equation}  
For other two-qp excitations the relations 
\begin{equation}
[T_\pm,\beta_{\mp\frac{1}{2},k}^+\beta_{\mp\frac{1}{2},l}^+]=
\beta_{\pm\frac{1}{2},k}^+\beta_{\pm\frac{1}{2},l}^+,
\end{equation}  
which follow from (\ref{com1qp}), imply 
\begin{equation}\label{tyeveneven}
[T_y,\frac{1}{\sqrt{2}}(\beta_{\frac{1}{2},k}^+\beta_{\frac{1}{2},l}^++
\beta_{-\frac{1}{2},k}^+\beta_{-\frac{1}{2},l}^+)]=0.
\end{equation}  
Only these linear combinations represent  $T=0$  two-qp excitations
in the even-even system. The combinations with the
minus sign must be discarded.  
The relations
\begin{eqnarray}
&[T_+,\beta_{\pm\frac{1}{2},k}^+\beta_{\mp\frac{1}{2},l}^+]=
\beta_{+\frac{1}{2},k}^+\beta_{+\frac{1}{2},l}^+, &\nonumber\\
&[T_-,\beta_{\pm\frac{1}{2},k}^+\beta_{\mp\frac{1}{2},l}^+]=
\beta_{-\frac{1}{2},k}^+\beta_{-\frac{1}{2},l}^+&
\end{eqnarray}  
imply that 
\begin{equation}\label{tyoddodd}
[T_y,\frac{1}{\sqrt{2}}(\beta_{-\frac{1}{2},k}^+\beta_{\frac{1}{2},l}^+-
\beta_{\frac{1}{2},k}^+\beta_{-\frac{1}{2},l}^+)]=0.
\end{equation}  
Only  these linear combinations  represent  $T=0$  two-qp excitations
in the odd - odd  system. The combinations with the
plus sign must  be discarded.

There is an alternative way to construct the $T_y=0$ configurations of
good number parity. We start with qp operators
$\alpha_{t_y,k}^+$ of good  $T_y$, which are given by  
the linear combinations  
\begin{eqnarray}
&\alpha_{\frac{1}{2},k}^+=\frac{1}{\sqrt{2}}(i\beta_{\frac{1}{2},k}^++
\beta_{-\frac{1}{2},k}^+),&\nonumber\\
&\alpha_{-\frac{1}{2},k}^+=\frac{1}{\sqrt{2}}(\beta_{\frac{1}{2},k}^++
i\beta_{-\frac{1}{2},k}^+).&\label{tytz1qp}
\end{eqnarray}  
Configurations of good number parity are constructed with the help 
of the relations
\begin{eqnarray}
&e^{-i\pi N}\alpha_{\pm\frac{1}{2},k}^+e^{i\pi N}=
\mp i\alpha_{\mp\frac{1}{2},k}^+,&\nonumber\\~~~
&e^{-i\pi Z}\alpha_{\pm\frac{1}{2},k}^+e^{i\pi Z}=
\pm i\alpha_{\mp\frac{1}{2},k}^+.&\label{numberparity}
\end{eqnarray}  
The $T_y=0$ excitation $\alpha_{\frac{1}{2},k}^+\alpha_{-\frac{1}{2},k}^+$
is odd under (\ref{numberparity}). Thus, it represents a $T=0$
state in the odd - odd nucleus. It is identical 
with $\beta_{\frac{1}{2},k}^+\beta_{-\frac{1}{2},k}^+$. Using
 (\ref{numberparity}) one sees that
\begin{equation}
e^{-i\pi N}\alpha_{\pm\frac{1}{2},k}^+\alpha_{\mp\frac{1}{2},l}^+e^{i\pi N}=
 \alpha_{\mp\frac{1}{2},k}^+\alpha_{\pm\frac{1}{2},l}^+
\end{equation}  
and that the $T_y=0$ excitations
 $\frac{1}{\sqrt{2}}(\alpha_{\frac{1}{2},k}^+
\alpha_{-\frac{1}{2},l}^+
\pm \alpha_{-\frac{1}{2},k}^+\alpha_{\frac{1}{2},l}^+)$
are even and odd under (\ref{numberparity}), respectively. They are identical
with the combinations appearing in the eqs. (\ref{tyeveneven}) and 
(\ref{tyoddodd}), as can be seen by means of direct evaluation using
eqs. (\ref{tytz1qp}).

An important $T_y$=0 four qp configurations in the even-even system is 
\begin{equation} 
\alpha_{\frac{1}{2},k}^+\alpha_{-\frac{1}{2},k}^+
\alpha_{\frac{1}{2},l}^+\alpha_{-\frac{1}{2},l}^+|0\rangle=
\beta_{\frac{1}{2},k}^+\beta_{-\frac{1}{2},k}^+
\beta_{\frac{1}{2},l}^+\beta_{-\frac{1}{2},l}^+|0\rangle,
\end{equation} 
 which appears in the double-alignment process.

It is noted that the number parity and $T_y$ do not commute,
\begin{equation}\label{numberparityty}
e^{-i\pi N}T_ye^{i\pi N}=e^{-i\pi Z}T_ye^{i\pi Z}=-T_y.
\end{equation}  
As a consequence of eq. (\ref{numberparityty}), only  the
eigenvalue $T_y=0$ is possible for states with good number parity.

\subsection{States with $T>0$}\label{isocrank.sec}

The $T>0$ bands are found by ``cranking in isospace'', employing the 
analogy between angular momentum and isospin. The ``frequency''
$\lambda_\tau$ is chosen such that \mbox{$\langle T_z\rangle =T$}. 
The corresponding configurations are interpreted as the 
states with maximal projection  $T_z=T$. The states with $T_z < T$ 
are the isobaric
analogs, which are generated by multiple application of $T_-$ on
 the configuration $|T,T_z=T\rangle$, generated by  cranking in isospace.
They have the same energy as the state $|T,T\rangle$.
 
Since the isospin symmetry is broken spontaneously, 
any infinitesimal value $\lambda_\tau$ in the  hamiltonian
(\ref{hfbh}) fixes the orientation of the pair-field 
perpendicular to the z-axis, i.e. $\Delta_{np}=0$. 
This explains, why in previous HFB studies  
solutions with $\Delta _{np}\not=0$ are found only for $ \lambda_\tau=0$
\cite{sandhu}.

The pn-pair-correlations appear  via the isospin symmetry.
 The state $|T,T\rangle$ can be interpreted as the pair-field being
oriented in the direction of no pn-correlation. The field executes
 zero-point oscillations around this equilibrium orientation that 
represent the  
pn-pair correlations, because any rotation away from the y-direction
introduces a pn-component in $\vec \Delta$. 
The amplitude of these  oscillations
quickly decreases with $T_z$ ( or  $\lambda_\tau$), which 
explains the  rapid decrease of the 
pn-pair correlations found in the good isospin calculations of ref.
\cite{engel}. The RPA theory of these oscillations has been worked out in
refs. \cite{camiz,ginocchio}. It is shown there that these oscillations 
contribute the term $\lambda_\tau/2$ to the energy. The appearance of this
term is well known for spatial rotation, which obeys the same
group $SU_2$. It corresponds to the 
familiar procedure to associate the angular momentum expectation value
calculated with the cranking wavefunctions to $I+1/2$
(e.g. \cite{bf79}) or  $\sqrt{I(I+1)}$ (e.g. \cite{ringschuck}). 
In the calculations we add $\lambda_\tau/2$ to the energy. 
It is likely that using the constraints,
$T+1/2=\langle T_z\rangle $ and $A=\langle N+Z\rangle $ to fix 
$\lambda_\tau$ and $\lambda$, 
will give very similar results, but we have not 
investigated this possibility.

Another way to approximately restore the symmetry  corresponds to
the high spin limit for spatial rotation.
For the state $|TT\rangle$, the probability amplitude for
the different orientations is given by the Wigner function  
${\cal D}^T_{TT}\propto (\cos \theta/2)^{2T}$. It is seen that the 
amplitude of the pn-field   ($\theta=\pi/2$) goes rapidly down.
For states $T_z<T$ the amplitude is ${\cal D}^T_{T_zT}$, which has large
contributions of the pn-field if $T_z$ is small.     

The excitation spectrum can be constructed in the familiar way by
quasiproton and quasineutron excitations, respecting the particle 
number parity.
Unlike for the $T=0$ states, the isospin symmetry does not exclude
configurations. The reason is that the probability for $\theta=\pi$, which
 corresponds to an exchange of a proton by a neutron, is 
zero in contrast to the case $T=0$. Nevertheless, for small $T_z$
the amplitude of the oscillations of the pair-field are substantial. This
will lead to interactions between the  quasiproton and
quasineutron states,  which will be discussed in section \ref{pncorr.sec}.

\subsection{Cranked Shell Model}\label{csm.sec}

In order to simplify the analysis of the excitation spectrum we shall
employ the CSM approximation \cite{bf79}.
It consists in solving the HFB equations for the ground-state configuration
at $\omega=0$. The states with finite angular momentum  are calculated from  
the qp routhians, which are the eigenvalues $e'_i$ of the  routhian
(\ref{hfbh}) keeping the mean-fields $\Gamma$ and $\Delta$ 
fixed to their values at $\omega=0$.  The energy of the configuration with the qps $\{ i, j, ...\} $
excited, is calculated as 
\begin{equation}
E'(\omega)=E'_o+e'_i(\omega)+e'_j(\omega)+....,
\end{equation}
where $E'_o$ is the energy of the reference configuration, the choice of which
is discussed in several reviews \cite{devoigt,bg} in detail. In the present
analysis,
it is calculated as the  HFB energy with the 
  $\omega =0 $ mean-field, i. e. 
\begin{equation} \label{e0csm}
E_o'= 
Tr [ (h'(\omega=0)+1/2 \Gamma)\rho+
1/2 \Delta t^\dagger ]
+\lambda_\tau /2.
\end{equation}
The fields $\Gamma$ and $\Delta$ are the ones found from
selfconsistency at $\omega=0$ and  $\Delta$ is taken in y-direction.
The density matrix $\rho$ and the pair-tensor $t$ are constructed 
from the eigenfunctions of the qp hamiltonian (\ref{hfbh}) using Eqs. (8)
and (9). The last term is the above discussed correction for the conservation
of isospin. 
                                                        
\section{Comparison of the cranked shell model with the exact
results}\label{csmsm.sec}

\begin{figure}[t]
\vspace*{-3cm}
\mbox{\psfig{file=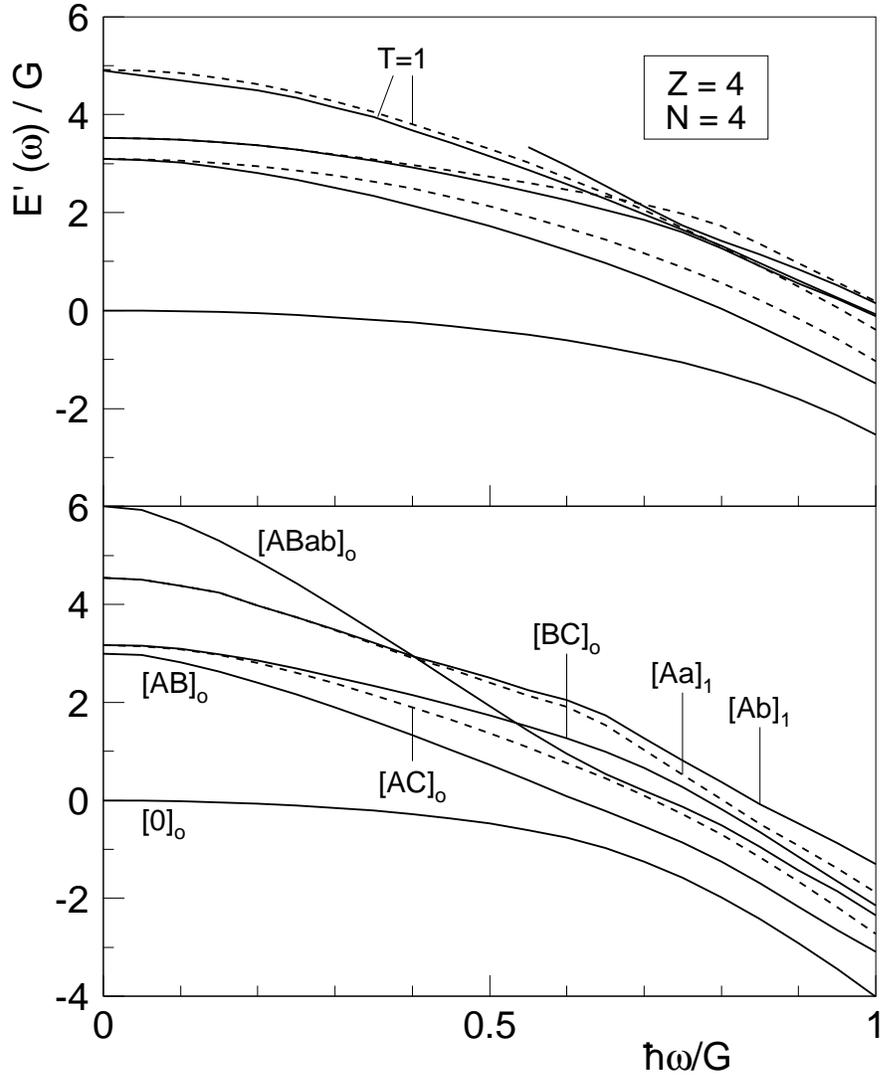,width=14cm}}
\caption{\label{e44.fig}
Total routhians for the $(Z=N=4)$ system. The upper panel shows
the SM results and the lower the CSM approximation.
Full lines correspond to even-spins and dashed as well as dashed
dotted ones to odd spins. 
The labeling of the quasiparticle configurations is 
explained in the text.   }
\end{figure}

We have calculated the exact energies by diagonalizing the SM
hamiltonian (\ref{H}) for $(Z+N=3+3)$, (3+4) and (4+4) particles in 
the $f_{7/2}$
shell. The results are shown in the upper panels of figs.
\ref{e34.fig} - \ref{e44.fig}.
The states are classified with respect to 
the isospin and the signature. We have solved the HFB equations
selfconsistently for (4+4) at $\omega=0$. The fields $\Gamma$ and $\Delta_{nn}=\Delta_{pp}$
determined thus are kept fixed for all other values of $\omega$.
They are also used to describe the (3+3) and (3+4) systems, for which only
$\lambda$ and $\lambda_\tau$ are adjusted 
to have $\langle  N \rangle  = N$ and $\langle  Z \rangle  = Z$ at $\omega=0$.

\begin{figure}[t]
\mbox{\psfig{file=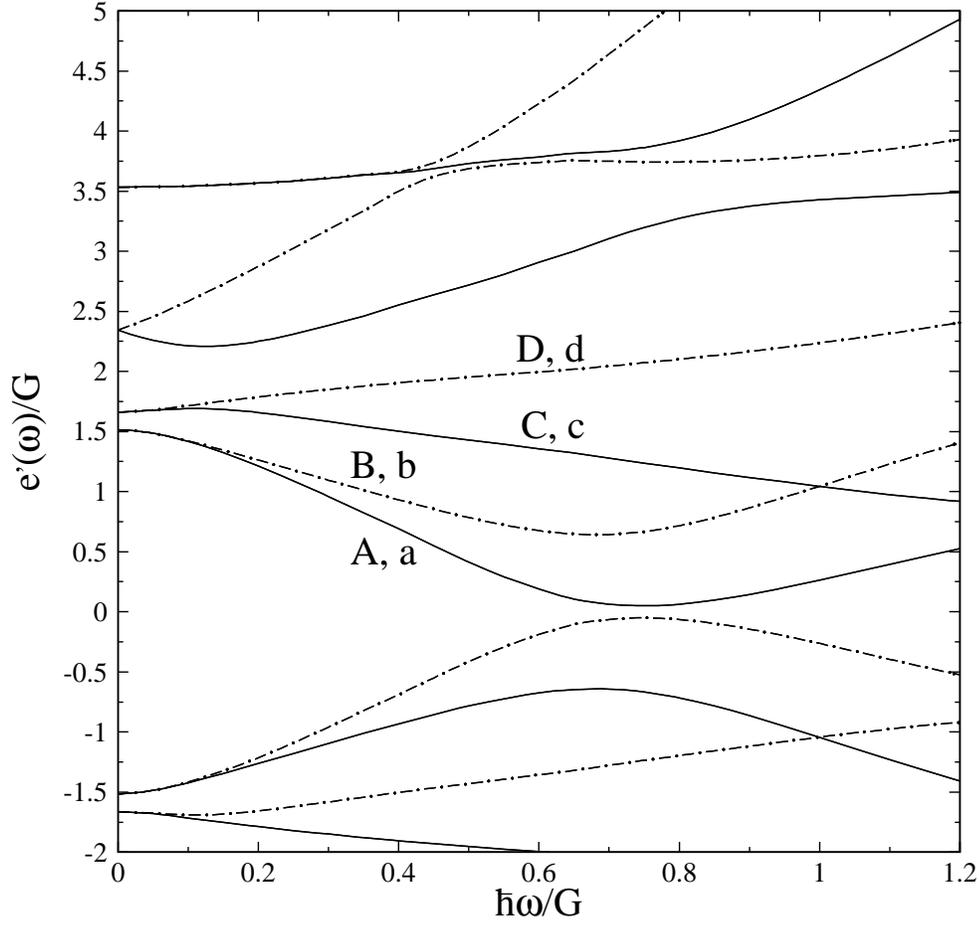,width=14cm}}
\caption{\label{qp.fig}
Quasiparticles in the $f_{7/2}$ shell as function of the rotational
frequency $\omega$. The chemical potential corresponds to a half filled
shell $\langle Z\rangle =\langle N\rangle =4$. The mean-field is kept fixed to the values
calculated  by solving the HFB equations (\protect \ref{hfb}) for 
$\omega=0$.  Full drawn and 
dashed dotted lines denote the favored and unfavored signature ($\alpha =-1/2$
and 1/2 for $f_{7/2}$), respectively.}
\end{figure}

Fig. \ref{qp.fig} shows the quasiparticle routhians $e'_i(\omega)$
 for the $(4+4)$ system. 
All the quasiparticle 
routhians are  two-fold degenerated, corresponding to a quasiproton
and a quasineutron , which are labeled, respectively, by a, b, c, ...
and A, B, C, ..., adopting  the popular CSM letter convention.
The degeneracy is lifted for $\lambda_\tau\not=0$. 

\subsection{Zero-quasiparticle configuration}\label{f720.sec}

\begin{figure}[t]
\mbox{\psfig{file=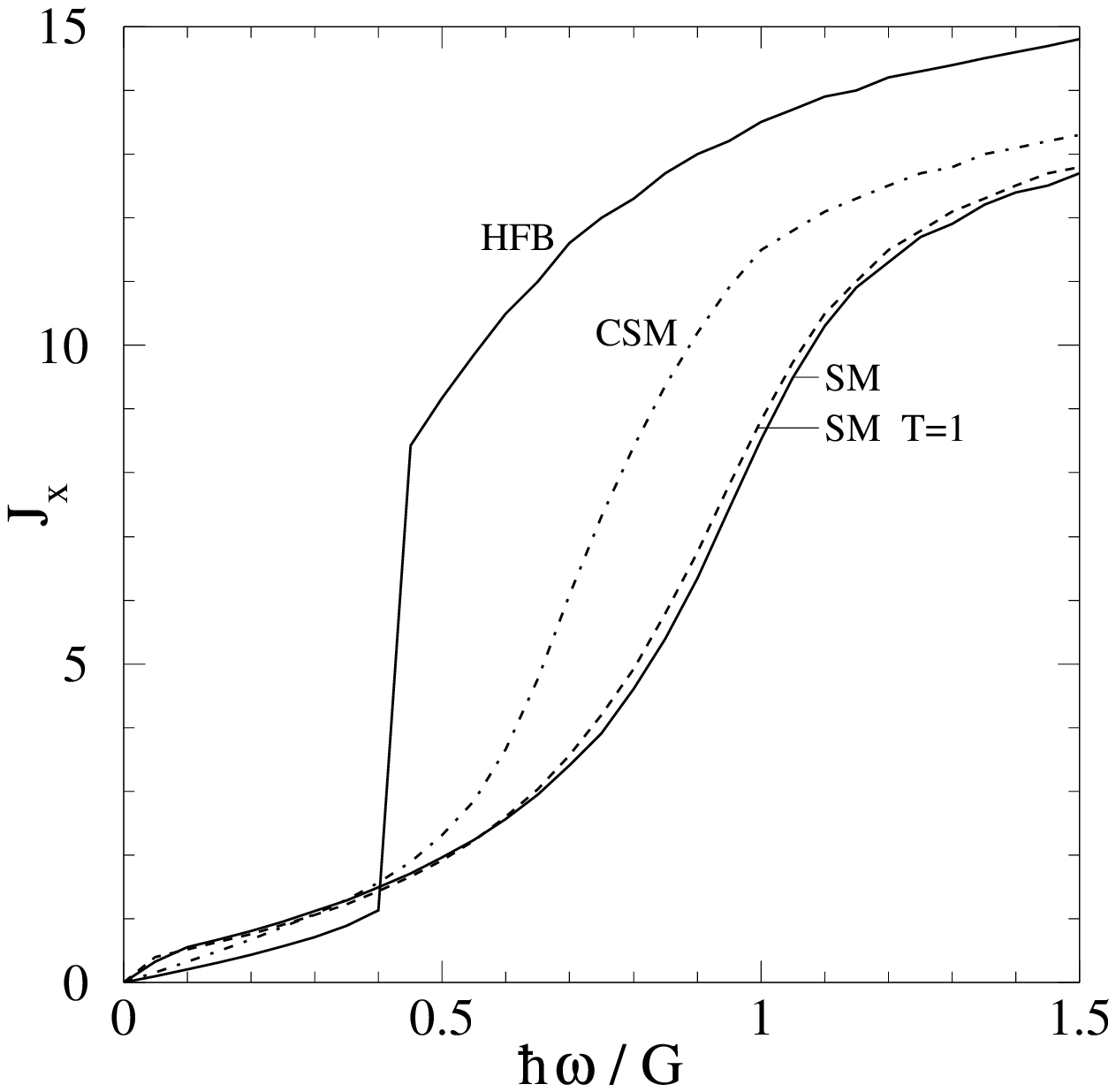,width=14cm}}
\caption{\label{jx.fig}
Angular momentum expectation value $\langle J_x\rangle $ for the yrast-band in 
the $(Z=N=4)$ system. The full SM result is denoted by SM, 
the  SM result with a modified two-body interaction leaving out 
the $T=0$ components 
of  the $\delta$-force by SM T=1, the fully selfconsistent HFB calculation
 by HFB and
the CSM approximation by CSM.}

\end{figure}

The configurations are constructed by the standard qp occupation scheme,
described in ref. \cite{bf79}. 
The vacuum [0] corresponds to all negative qp orbitals filled.  
It has signature
$\alpha =0$, even-$N$ and -$Z$ and $T_y=0$. It represents the even-spin $T=0$ 
yrast-band of the $(N=Z=4)$  system. The AB-crossing at $\omega=0.6$
corresponds to the {\em simultaneous} alignment of a proton- and a neutron-
pair (because the routhians are degenerate). Since each $f_{7/2}$
pair carries 6 units of alignment, the total gain amounts to 12 units.
The double-alignment
as a specific feature of  even-even  $N=Z$ nuclei 
has first pointed out in ref. \cite{she90}.
Fig. \ref{jx.fig}  demonstrates that the CSM approximation 
describes the double-alignment fairly well, the
crossing frequency being somewhat under estimated. This discrepancy will be
commented in section \ref{pncorr.sec}.
The good agreement between the exact SM calculations
and the CSM has first been noticed in ref. \cite{pnr89} for the case
of one kind of particles in a $j$-shell.
It is important to take into account of the renormalization 
of the single particle levels by the interaction
($\Gamma$ in eq. (\ref{hfbh})) to obtain the  agreement.
As seen in fig. \ref{jx.fig}, the calculation 
where the fields are determined selfconsistently for
all $\omega$ shows a much to early alignment. A similar result has been
found in ref. \cite{pnr89} for one kind of particles in a $j$-shell.
Thus, for the model case of a single $j$-shell, the CSM approximation mocks up
some of the corrections to the mean-field approxiamtion. 

\subsection{One-quasiparticle configurations}\label{f72oa.sect}

\begin{figure}[t]
\vspace*{-3cm}
\mbox{\psfig{file=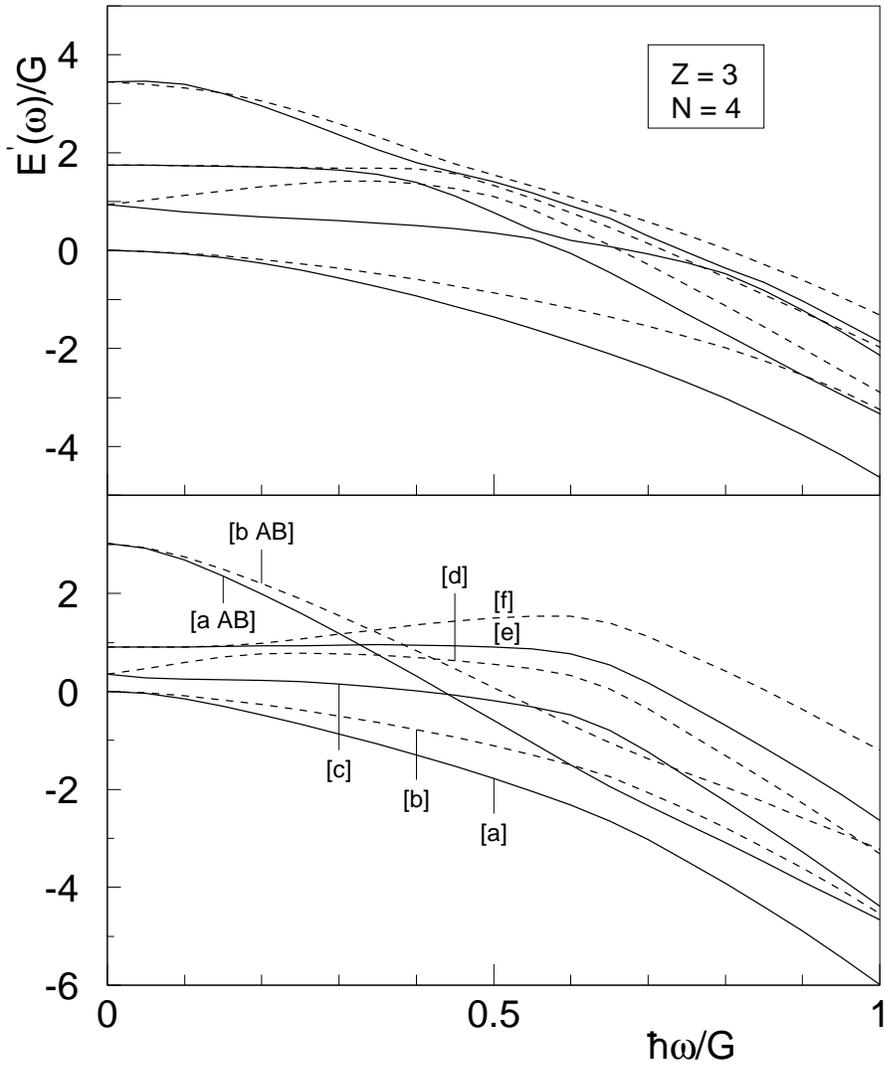,width=14cm}}
\caption{\label{e34.fig}
Total routhians for the $(Z=3, N=4)$ system. The upper panel shows
the SM results and the lower the CSM approximation.
Full lines correspond to signature -1/2 and dashed ones to 1/2. 
The labeling of the quasiparticle configurations is 
explained in the text.}
\end{figure}

The simplest configurations are generated by exciting one-quasiparticle 
to the  lowest routhians.
They correspond to the $T=1/2$ bands in the odd-A nuclei. Fig. 
 \ref{e34.fig} shows the case $(Z=3$,$N=4)$.
The lowest bands correspond to  the one-quasiproton configurations 
 [a], [b], [c], [d]. Their excitation energies agree rather  well 
with the exact SM calculation for the lowest bands, which
have, as expected, $T=1/2$ and signature 
$\alpha=-1/2,1/2,-1/2$ and 1/2, respectively.

All bands show a crossing with three qp bands that contain the pair [AB]
of aligned quasineutrons. In Fig. \ref{e34.fig}, only the configurations
[aAB] and [bAB] are included before the crossing.
It is seen that the crossings  occur systematically 
at a higher frequency in the SM calculation than in the CSM. 
This effect will be discussed in section \ref{pncorr.sec}.

\subsection{Two-quasiparticle excitations in the odd-odd system}
\label{f72oo.sec}

\begin{figure}[t]
\vspace*{-3cm}
\mbox{\psfig{file=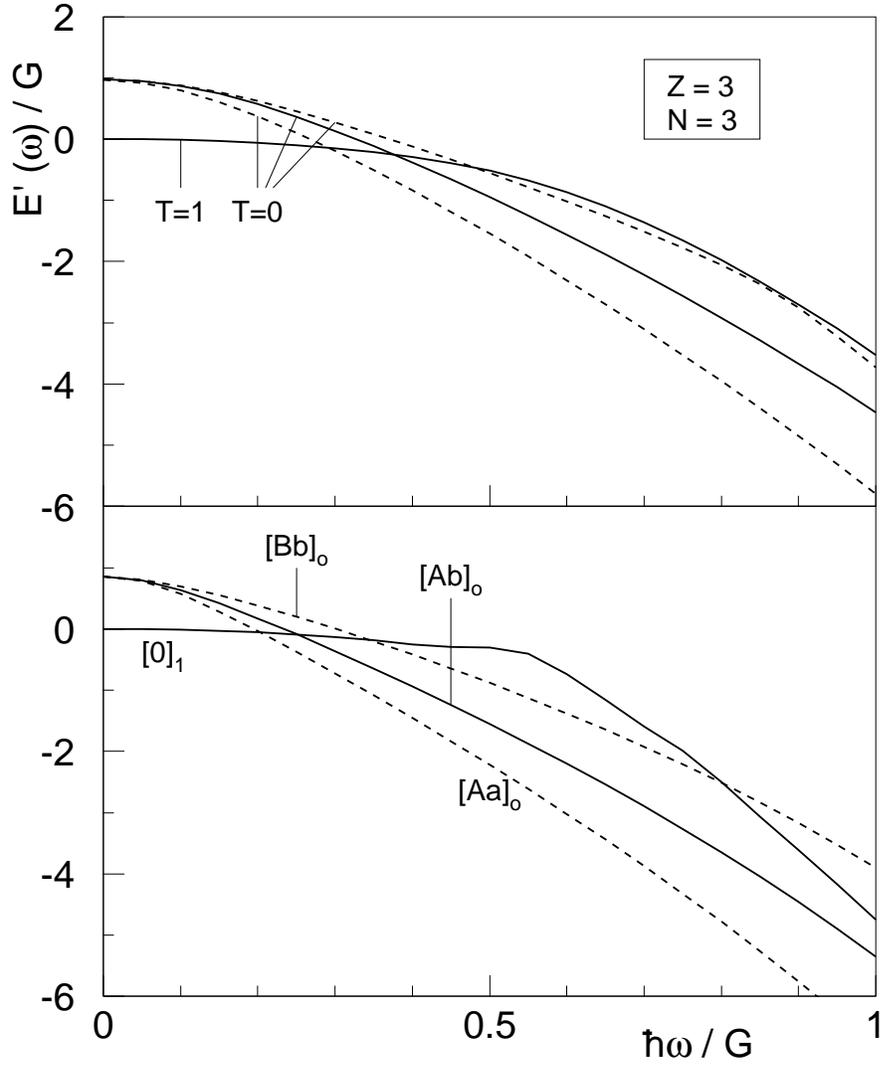,width=14cm}}
\caption{\label{e33.fig}
Total routhians for the $(Z=N=3)$ system. The upper panel shows
the SM results and the lower the CSM approximation.
Full lines correspond to even-spins and dashed ones to odd spins. 
The labeling of the quasiparticle configurations is 
explained in the text. }
\end{figure}

The lowest two-qp excitation is generated by putting one-quasiproton and
one-quasineutron on the lowest  routhian. We denote
this configuration by $[A,a]_0$. As discussed in section \ref{t=0.sec},
it has $T_y=0$ and thus correspond to a $T=0$ band. 
The subscript indicates the isospin $T$ of the configuration.
The total signature is $\alpha=1$ and corresponds to an odd-spin band.
The particle numbers
$N$ and $Z$ must be odd, because exciting one-quasineutron 
 changes $N$ from even to 
odd or from odd to even and the same holds for the quasiprotons. 
Thus  $[A,a]$
 is the lowest $T=0$ odd-spin band in the 
odd-odd $N=Z$ system. Fig. \ref{e33.fig}
shows the CSM estimate for this band, which
is obtained  choosing $\lambda$ such 
that $\langle N\rangle =\langle Z\rangle =3$
but using the $\omega=0$ mean-field parameters calculated self consistently 
for $\langle N\rangle =\langle Z\rangle =4$. It is important to note that
the vacuum $[0]_0$ obtained for the same
 $\lambda$ does not represent a physical
state, because it corresponds to even-$N$ and even-$Z$.
The configuration $[B,b]$ is the second odd-spin $T=0$ band and
$[A,b]$ the first even-spin $T=0$ band in the odd-odd system.      
As discussed in section \ref{t=0.sec}, the configuration $[a,B]$ 
does not generate a new state, because
 isospin $T=T_y=0$  corresponds to the superposition $([A,b]-[a,B])/\sqrt{2}$. 
To keep the notation simple, we label the configuration as $[A,b]$. But it is
understood that the superposition is meant.
These are the three lowest  $T=0$ bands. 

The lowest $T=1$ band is found by ``cranking in isospace''. We calculate the
isobaric analog state \mbox{$|T=1,T_z=1\rangle$} by adjusting $\lambda$ and
 $\lambda_\tau$ such that\mbox {$\langle Z\rangle =2$} and 
\mbox{$\langle N\rangle =4$}.
According to CSM ideology, the other mean-field
parameters are kept to values found 
for the  $\langle N\rangle =\langle Z\rangle =4$ system. The qp
spectrum looks like fig. \ref{qp.fig}, except that the quasineutron 
routhians (A, B, ...) and
quasiproton routhians (a, b, ...) are no longer  degenerate. The vacuum,
 which we denote by 
$[0]_1$, starts as the g-configuration, being crossed first by $[ab]$ and
then by $[AB]$. It represents the even-spin $T=T_z=1$
yrast-band of the even-even $Z=N-2$ system. The configuration  $[0]_1$
appears  as  isobaric analog band in the odd-odd $Z=N$ system,
where it represents the lowest  even-spin $(T=1, T_z=0)$ band.

The comparison with the SM calculation  in fig. \ref{e33.fig} 
demonstrates that this simple procedure of ``cranking in isospace'' 
reproduces  well the position of the $T=1$ even-spin band 
relative to the three lowest $T=0$ bands, the relative position of
which is also well
reproduced by the CSM.  The appearance of the 
$T=1$ even-spin band below  the $T=0$ bands
is a specific feature of the $Z=N$ system.( In   odd-odd nuclei
with $N\gg Z$
all  bands start with an energy larger than $2\Delta$.) Its low-energy 
for  $\omega=0$ has the consequence that 
the  $T=1$ even-spin band is crossed by
the aligned odd-spin $T=0$ band. This crossing has been observed in 
$^{74}$Rb \cite{rb74}. 
The similar energy of the $T=1$ and $T=0$ states at $\omega=0$ appears
 as a cancelation between the pair-gap and the ``iso-rotational''
energy. Relative to the $\lambda_\tau=0$ qp vacuum, the configuration
$[Aa]_0$ is shifted by $2\Delta$. The configuration $[0]_1$ is shifted
by $T(T+1)/2{\cal J}=1/{\cal J}=\lambda_\tau/\langle T_z\rangle $. Both quantities are 
nearly equal. This is not a special feature of our $j$-SM, but    
a quite general phenomenon, as discussed in ref. \cite{vogel}.

The CSM reproduces SM value for 
the energy difference between the $T=0$ and $T=1$ at $\omega=0$ rather
well. The correction 
$\lambda_\tau /2=0.88$ in the ground state energy (\ref{e0csm})
 considerably improves the agreement. The crossing frequency
between the $T=1$ and $T=0$ bands is somewhat underestimated by the CSM, 
 because it  overestimates the alignment of
the odd-spin band ( for $\omega= 0.3 G$ the CSM gives $J_x=5.7$
and the  SM $J_x=4.7$ ).

\subsection{Two-quasiparticle excitations in the even-even system}
\label{f72ee.sec}

Fig. \ref{e44.fig} shows the energies of the lowest bands in the 
system (4+4). Most of them have $T=0$. In oder to have 
even-$N$ and -$Z$ one must excite
two quasiprotons or two quasineutrons. The
lowest excitation is $[AB]$ which has signature $\alpha=0$ and
is degenerate with [ab]. Out of these two, only
the combination $([A,B]+[a,b])/\sqrt{2}$ has 
$T_y=0$ and represents a $T=0$ even-spin band. In order to keep the notation
simple, we denote this band by $[AB]_0$.  Again it is
understood that the superposition is meant. The next two $T=0$
configurations are the $\alpha=1$  (odd-spin) band $[AC]_0$ and the  
$\alpha=1$ (even-spin) band  $[BC]_0$. These bands are crossed by the
$T=0$  four qp band $[abAB]_0$, which causes the double-alignment in the yrast-
line. This structure is clearly correlated with the sequence of $T=0$ bands
in the SM calculation. The CSM underestimates the excitation
energy of $[AB]_0$ and $[abAB]_0$, which has the consequence that the alignment
in the yrast-band comes too early.

There is a discrepancy at low-$\omega$. In the CSM, the $[AB]$ configuration
is separated from $[AC]$ and $[BC]$, which become a signature doublet and
degenerate with the next doublet $[AD]$ and $[BD]$.
For the SM, the lowest bands at small $\omega$ are two signature doublets
with a finite energy difference. We have not analyzed this discrepancy.
It is possible that for small $\omega$ the assumption of stable rotation 
about the x-axis is violated and the system transiates to a tilted
rotational axis \cite{tac}. This would result in substantial changes
of the spectrum.

The lowest $T=1$ configurations are obtained by ``cranking in isospace''.
In the CSM spirit, we take the (4+4) mean-field and adjust $\lambda$ and
 $\lambda_\tau$ such that $\langle Z\rangle =3$ and $\langle N\rangle =5$ and include the correction 
 $\lambda_\tau/2=1.01$ in $E_o$. The configurations must have odd $N$ and
odd $Z$, because the isobaric analog state $|11\rangle$ belongs to the
odd-odd $(3+5)$ system. The lowest configuration is the 
$\alpha=1$ configuration
$[aA]$ and the next the $\alpha=0$  configuration $[aB]$. They represent the
odd-spin band $[aB]_1$ and even-spin band $[aB]_1$, respectively.
They are nearly degenerate, only at high $\omega$ there appears some
signature splitting, favoring the odd-spin band. 
In the SM calculation there is an odd- and an even-spin
band with $T=1$ at about the right energy. They show a small signature
splitting. 
Hence, the CSM also accounts well for the lowest excited states in
the even-even system.

\section{Cranked shell model analysis of nuclei with $A\approx 72$}

The analysis of realistic nuclei 
mainly follows the CSM procedure as described in ref. \cite{bf79}. We 
will point out the modifications ensued by the isospin conservation.
The modified harmonic 
oscillator potential  with the standard set of Nilsson parameters
as given in ref. \cite{nilsson} and 
the deformations $\varepsilon=0.3, ~\gamma=0$
and $\varepsilon_4=0$ and
the monopole pair-fields  with $\Delta_p=\Delta_n=1.1 MeV$,
corresponding  to about 80\% 
of the experimental even-odd mass differences, are used. 
Calculations of equilibrium
deformations within the shell-correction method \cite{galeriu,kr74,rb75}
show that the deformations at 
low-frequency are characterized by a coexistence of prolate and oblate shapes
and softness with respect to the triaxiality parameter $\gamma$. 
At larger frequencies all nuclei tend to take on a near prolate shape
with $\varepsilon = 0.3-0.4$. Since we are interested in the 
qualitative structure of the yrast-spectra of the $N \approx Z$ nuclei 
at large spin, the deformation is kept fixed at $\varepsilon=0.3$.

The experimental routhians are calculated along the lines of 
ref. \cite{bf79} as
\begin{eqnarray}\label{exprouth}
\omega=\frac{1}{2}(E(I)-E(I-2))
\sqrt{1-(\frac{K}{I-\frac{1}{2}})^2},\\
E'=\frac{1}{2}(E(I)+E(I-2))-(I-\frac{1}{2})\omega.
\end{eqnarray}
The frequency has the meaning of the 1-component of angular velocity.
The symmetry axis is 3 and  and $K$ the  component of the angular momentum
along this axis,
which is kept constant.\footnote{ Eq. \ref{exprouth} represent  
a slight modification of the expressions given in \cite{bf79}~ and is
more accurate near the bandhead \cite{meng}.}
A common Harris reference
\begin{equation}
E'_o=-\frac{\omega^2}{2}{\cal J}_o-\frac{\omega^4}{4}{\cal J}_1+
\frac{1}{8{\cal J}_o}
\end{equation}
 is subtracted from all the
experimental routhians. The parameters ${\cal J}_o=13 MeV^{-1}$ and 
${\cal J}_1=8 MeV^{-3}$
are fitted to the experimental yrast-energies in $^{72}$Kr. Likewise,
a common Harris reference is also subtracted from the calculated 
routhians. The  parameters ${\cal J}_o=8.4 MeV^{-1}$ and 
${\cal J}_1= 25 MeV^{-3} $ are fitted to the 
calculated routhian of the ground (g-) band in $^{72}$Kr.  

\begin{figure}[t]
\mbox{\psfig{file=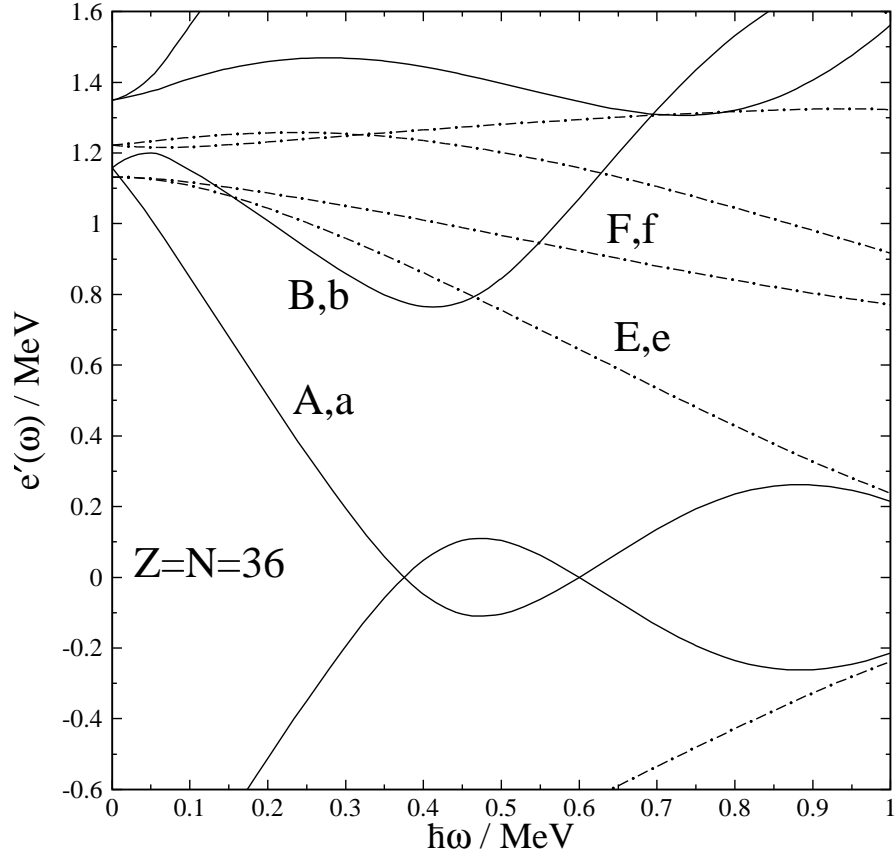,width=14cm,angle=-90}}
\caption{\label{qp72.fig}
Quasiparticles for  $(N=Z=36)$  as function of the rotational
frequency $\omega$.  The mean-field is
the modified oscillator with the deformations $\varepsilon=0.3$,
$\varepsilon_4=0$ and $\gamma=0$ and $\Delta_n=\Delta_p=1.1 MeV$.
The diagram is relevant for both protons and neutrons. 
Full drawn and 
dashed dotted lines denote positve and negative parity, respectively.
The signature is indicated by the letters: $\alpha=1/2$ for A,E and
$\alpha=-1/2$ for B,F. }
\end{figure}

Fig. \ref{qp72.fig} shows the quasineutron routhians for $N=36$.
The quasiproton routhians are nearly identical (The slight deviations  are due
to differences of the Nilsson potential for protons and neutrons.)
The standard letter coding is used to label the qp routhians. The use of
A, a, ...  indicates that the diagram \ref{qp72.fig} is relevant for both
neutrons and protons.

\subsection{$^{73}$Kr}\label{kr73.sec}

\begin{figure}[t]
\vspace*{-3cm}
\mbox{\psfig{file=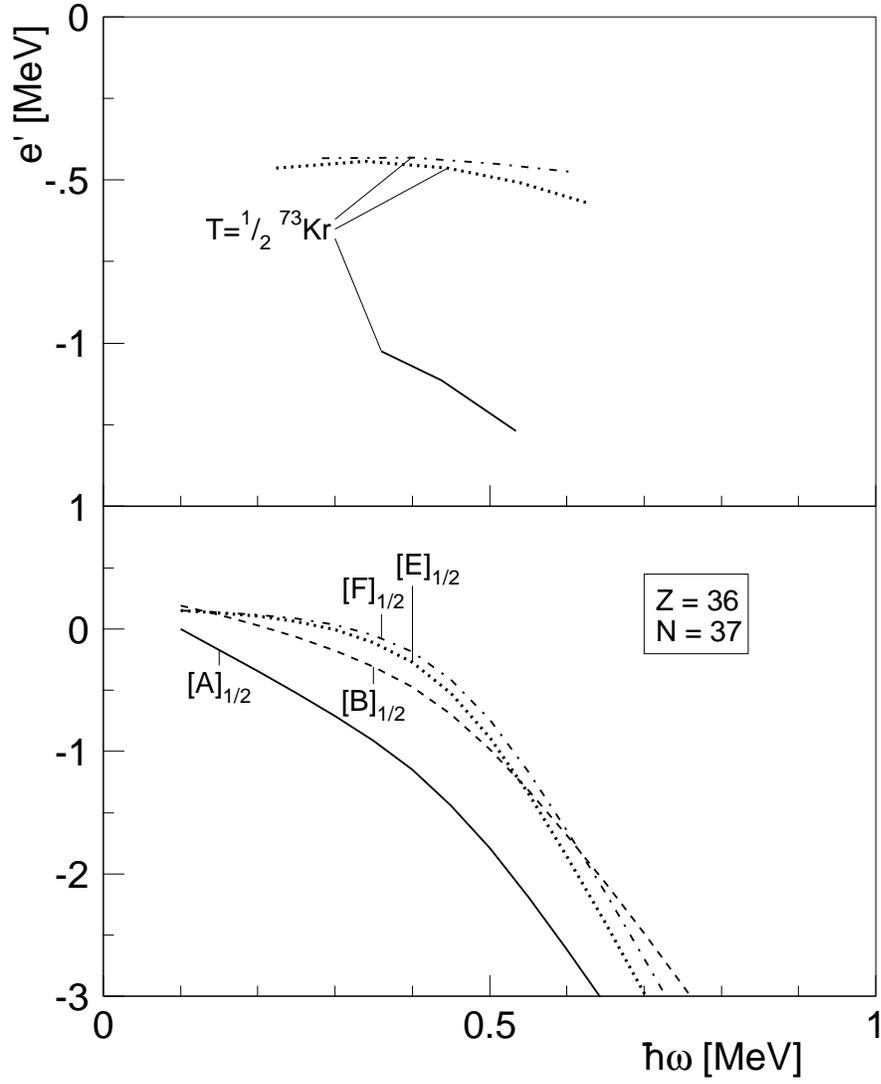,width=14cm}}
\caption{\label{kr73.fig}
Total routhians for the $^{73}_{36}$Kr$_{37}$. The upper panel shows
the experimental routhians \protect\cite{kr73}
 and the lower the CSM approximation.
The parity and signature assignments $(\pi,\alpha)$ are:
Full lines (+,1/2), dashed (+,-1/2), dashed dotted (-,1/2) and
dotted (-,-1/2). A Harris reference is subtracted.
   }
\end{figure}

Fig. \ref{kr73.fig} shows the experimental and CSM routhians for 
$^{73}_{36}$Kr$_{37}$. The low-frequency part consists of the one-
quasineutron configurations [A], [B], [E], [F], .... The calculated
routhians slightly deviate from fig. \ref{qp72.fig}, because $\lambda_\tau$
is adjusted to have $\langle N\rangle =37$. The relative position of the trajectories
A, E, F are reasonably well reproduced. At $\omega=0.5 MeV$ the calculated 
routhians bend downwards as a consequence of $g_{9/2}$ alignments. In the case
of [A] and [B], it is the proton s-band [ab] that crosses. In the case
of [E] and [F], it is the double s-band [ABab]. The 
negative parity bands become yrast because  the neutron alignment 
[AB] is blocked in [A] and [B].  The CSM under estimates the frequency of
these band crossings. There are the pn-correlations which are not
included in the CSM, that systematically delay the crossing. The systematic investigation of these shifts in ref. \cite{fs} indicates that for [A] 
the crossing 
should be similarly delayed as in $^{72}$Kr, where the double s-bands
crosses at $\omega=0.75$. It will be interesting to see where the double
s-band crosses  in the case of [E] and [F]. If it will cross
earlier than in $^{72}$Kr, this would be evidence that the pairing channel
is important for the correlations, because substantial blocking is expected 
due to E and F. The absence of the alignments in
the data seems to indicate a substantial delay for
 both parities, but definite conclusions cannot be drawn before
the alignments have been observed.

\subsection{$^{74}$Rb}\label{rb74.sec}

\begin{figure}[t]
\vspace*{-3cm}
\mbox{\psfig{file=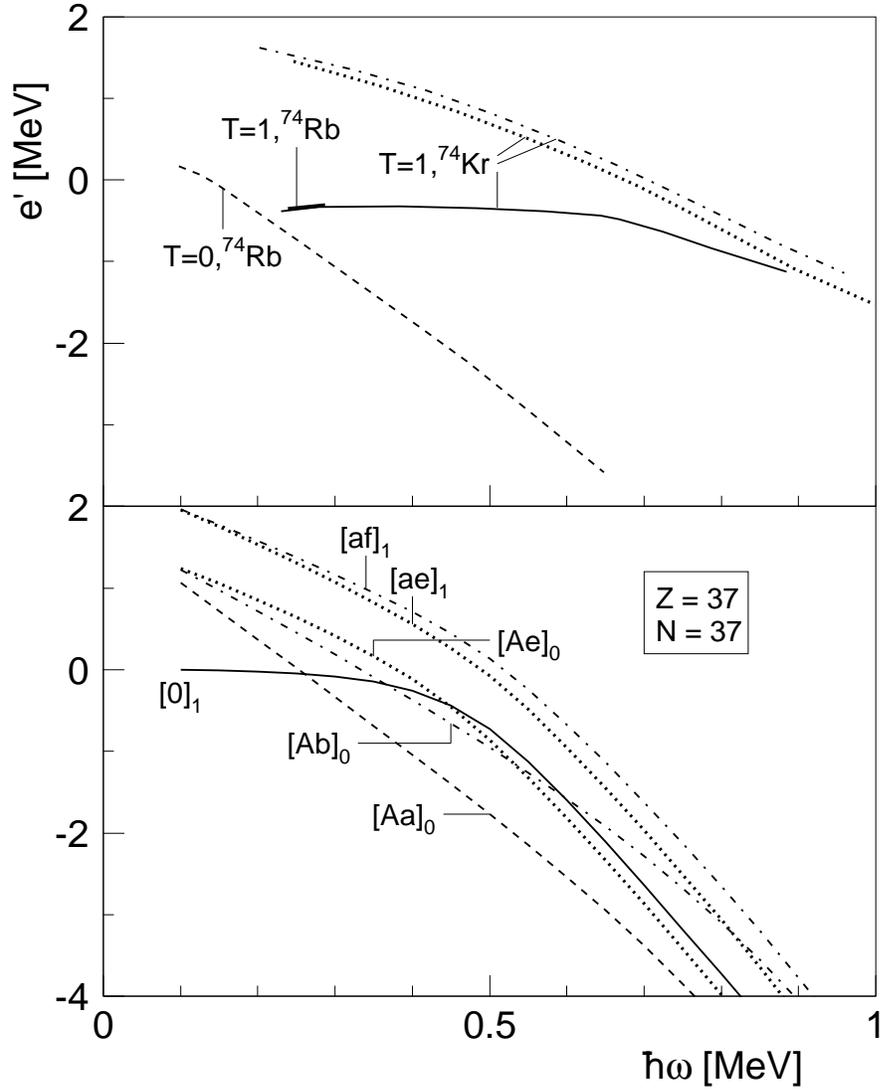,width=14cm}}
\caption{\label{rb74.fig}
Total routhians for  $^{74}_{37}$Rb$_{37}$. The upper panel shows
the experimental routhians \protect\cite{rb74}
 and the lower the CSM approximation. For $T=1$ also the isobaric
analog $T_z=1$ bands in $^{74}_{36}$Kr$_{38}$  \protect\cite{kr74}
are shown. 
The parity and signature assignments $(\pi,\alpha)$ are:
Full lines (+,0), dashed (+,1), dashed dotted (-,0) and
dotted (-,1). A Harris reference is subtracted.
   }

\end{figure}

Fig. \ref{rb74.fig} displays the spectrum of  $^{74}_{37}$Rb$_{37}$.
The upper panel also shows   the $T_z=1$ bands measured in
$^{74}_{36}$Kr$_{38}$. They are  isobaric analog to the $T=1$ bands in 
$^{74}$Rb and should give a good estimates of these  bands.
 Since the ground states belong to an isobaric triplet, 
we set the energy $E_o$ of $^{74}$Kr equal to the one
of $^{74}$Rb. As seen, the routhians of the $4^+ \rightarrow 2^+$
and  $2^+ \rightarrow 0^+$ transitions in both  nuclei  are nearly identical.
The experimental verification of the expectation that 
  $T_z=1$ bands are rather good estimates of the $T_z=0$
bands, which are not yet measured, 
and the study of the fine differences 
will shed new light on the breaking  of the isospin conservation
by the Coulomb potential. 

The lowest $T=0$ configurations are generated by exciting a 
quasiproton  and a quasineutron.
The first is the positive parity odd-spin band $[Aa]_0$.
 Next,
$[Ab]_0$ and $[Ae]_0$  are expected. 
 As discussed in section \ref{t=0.sec},
the condition $T_y=0$ permits only one linear combination 
 of the two excitations, obtained by exchanging the quasi
proton with the quasineutron, which we arbitarily label by
only one of the terms in order to keep the notation simple.
Analog to $^{73}$Kr, $[Ae]_0$ crosses
$[Ab]_0$, because b blocks the ab - alignment, but e does not.

The lowest $T=1$ bands are generated by cranking in isospace.
Thus, $\lambda_p$ and $\lambda_n$ are fixed to have $\langle Z\rangle =36$
and $\langle N\rangle =38$ at $\omega=0$ and kept fixed for the other $\omega$ values.
 The isospin correction energy $\lambda_\tau/2$,
which is in this case 0.39$MeV$ is included in $E_o$. The lowest
band is the vacuum $[0]_1$. It is crossed by the $T=0$ band $[Aa]_0$,
which has a large alignment. (In the CSM calculation it is 
about 7  and in experiment
with the chosen reference it is changing from 3 at the bottom
to  7 for the highest observed transition). The crossing frequency 
is fairly well reproduced. Thus it seems, that this crossing is 
a phenomenon belonging to the realm of $t=1$ pair-correlations. 
As already discussed for the $f_{7/2}$ model case, the close
energies of the lowest $T=0$ and $T=1$ states result from the
near cancelation of two large energies: The pair-energy $2\Delta$,
by which the $T=0$ state with two-qp character is shifted and the 
iso-rotational energy $T(T+1)/2{\cal J}=\lambda_\tau/\langle T_z\rangle $ 
by which
the  $T=0$ state with zero-qp character is shifted. The latter is
somewhat smaller than the former (numbers are given
in subsection \ref{kr72.sec}). An accurate estimate of the 
crossing frequency cannot be expected from our the CSM calculation, 
 because  it simply assumes
that $\Delta$ is 80$\%$ of  experimental even-odd mass difference, which 
is pretty rough.  The estimate of ${\cal J}$ ($\propto$ the 
level density) is also  rough, because the  common
deformation of $\varepsilon =0.3$ and $\gamma=0$ is assumed
for all the considered nuclei. Optimizing the shape  will reduce the level
density near the Fermi surface and, thus, make ${\cal J}$ smaller
and push up the $T=1$ band. 
   
The transition $(3^+)\rightarrow (2^+)$ observed in $^{74}$Rb 
is of M1 type, if the tentative spin and parity assignments are correct.
Since the M1-transition operator is predominantly isovector,
it favors transitions $T=0 \rightarrow T=1$. A measurement of the lifetime 
would be quite interesting, because the  difference in 
aligned angular momentum 
(about 3 for the frequency of the observed transition)
makes the $B(M1)$ value sensitive to isospin impurities.

\subsection{$^{72}$Kr}\label{kr72.sec}

\begin{figure}[t]
\vspace*{-3cm}
\mbox{\psfig{file=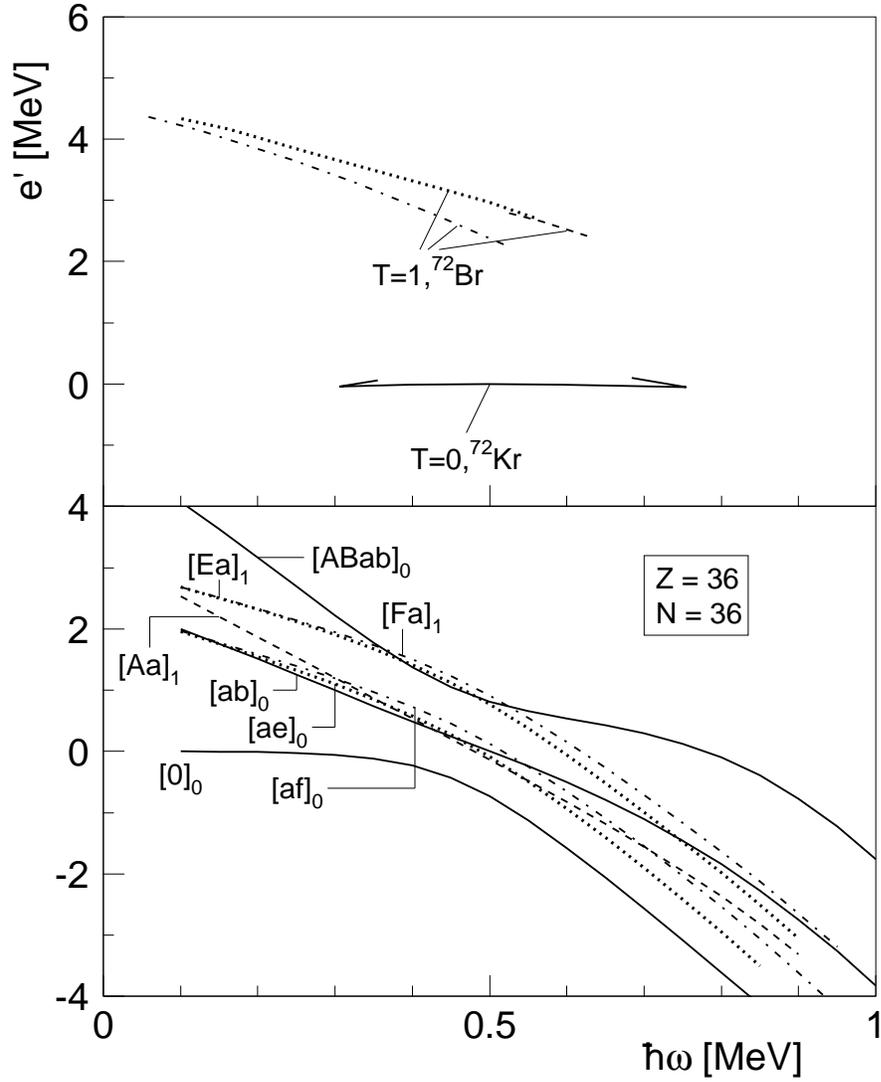,width=14cm}}
\caption{\label{kr72.fig}
Total routhians for the $^{72}_{36}$Kr$_{36}$. The upper panel shows
the experimental routhians \protect\cite{kr72}
 and the lower the CSM approximation. For $T=1$ also the isobaric
analog $T_z=1$ bands in $^{72}_{35}$Br$_{37}$  \protect\cite{br72}
are shown. The text explains how the energy of the $T=1$ bands  relative
to the  energy of the $T=0$ ground state is fixed.
The parity and signature assignments $(\pi,\alpha)$ are:
Full lines (+,0), dashed (+,1), dashed dotted (-,0) and
dotted (-,1). A Harris reference is subtracted.
   }
\end{figure}

Fig. \ref{kr72.fig} displays the spectrum of  $^{72}_{36}$Kr$_{36}$.
The upper panel shows also  the $T_z=1$ bands measured in
$^{72}_{35}$Br$_{37}$. They are expected to be isobaric analog.
Since the ground state of $^{72}$Kr has $T=0$, the energy difference
between the lowest states of each isospin must be estimated.
We use the expressions discussed in ref. \cite{vogel},
\begin{equation}\label{esymm} 
E_{symm}=\frac{T(T+1)}{A}[134.4-203.6A^{-1/3}]MeV
\end{equation}
which can be considered as an phenomenological expression
for the iso-rotational energy,
fitting the experimental binding energies, and
\begin{eqnarray}\label{twodel} 
&\Delta=5.39A^{-1/3}MeV ~(N=Z+2),\nonumber&\\ &\Delta=6.24A^{-1/3}MeV~ (N=Z)&.
\end{eqnarray}
The ground state of $^{72}$Br is then at $E_{symm}+2\Delta$. The numbers for
$A=72$ are $E_{symm}=2.37 MeV$ and $2\Delta=2.59$. As discussed in sect.
\ref{rb74.sec}, 
the difference $(E_{symm}-2\Delta)$ gives the distance between the lowest
$T=1$ and $T=0$ states in the odd-odd nuclei. For $^{74}$Rb, one finds
$E_{symm}=2.32 MeV$ and $2\Delta=2.97MeV$. The difference of 0.65MeV
can be compared with the experimental difference of 0.57 MeV
between the intrinsic energies of the $T=1$ and 
$T=0$ states in $^{74}$Rb.\footnote{ The intrinsic energy of the
$T=1$ band is estimated assuming that the rotational     
energy is given by the strong coupling expression 
$(I(I+1)-K^2)/{\cal J}_{rot}$ with  $K=2$ and fitting ${\cal J}_{rot}$
to the $(3^+)\rightarrow (2^+)$ transition.}

The yrast band is the $T=0$ configuration $[0]_0$. It is crossed by
$[ABab]_0$. This double-alignment is observed in  the experimental
yrast sequence \cite{kr72} at a substantial higher frequency than in the CSM.
The next bands are  $[ab]_0$, $[ae]_0$ and $[af]_0$, where we use again the
 short hand notation for the  $T_y=0$ linear combinations.
The $f_{7/2}$ SM study suggests that the configuration
 $[ab]_0$ is predicted
too low by the CSM (c. f. section \ref{f72ee.sec}.   

The lowest $T=1$ configurations  are $[Aa]_1$, $[Ea]_1$ and $[Fa]_1$.
They are experimentally seen as the isobaric analog
 bands in $^{72}$Br. Using these energies and the above described
estimate for the relative energies of the $T=0$ and $T=1$ ground states
the ``experimental''  $T=1$ bands in $^{72}$Kr lie about 1 $MeV$ higher
than our CSM estimate. As already discussed in section \ref{rb74.sec},
the CSM estimates for $2\Delta$ and $E_{symm}$, which determine the relative
energy of the $T=0$ and $T=1$ bands are pretty rough in our CSM calculation.

\subsection{TRS calculation for $^{74}$Rb}\label{trs.sec}

The CSM assumptions of fixed deformation and pairing are too inaccurate
for the high frequency region of the considered nuclei. 
Here we present the spectrum of $^{74}$Rb as an example that the 
concepts suggested in this paper can be 
combined with more sophisticated mean-field calculations. 

The TRS
calculations presented in ref. \cite{kr74} describe rather well 
the rotational bands in $^{74}$Kr. The deformed 
Woods-Saxon potential with the "universal parameters" is used. 
Only pp- and nn- pairing is
considered, but in addition to the  monopole a quadrupole pair-field is
taken into account. For each configuration and frequency $\omega$,
 the deformation parameters are
individually optimized.   The details of the calculation are described in
ref.  \cite{kr74}. The calculations of  \cite{kr74} for the 
yrast sequence in $^{74}$Kr are used  for the configuration $[0]_1$ and
the results of an analogous TRS calculation \cite{rb74trs}
 for $^{74}$Rb are used for the configurations $[Aa]_0$ and $[Ae]_0$.
\footnote{The TRS calculations
have been carried out by W. Satula. The authors would like to express their
gratitude for making these results available to them.}
The relative energy of the $T=0$ and $T=1$ bands is calculated 
by setting at $\omega=0$ the energy difference between the
configurations $[0]$  in $N=38, ~Z=36$ and $N=37, ~Z=37$ equal to
the expression (\ref{esymm}) for the iso-rotational energy.                
The same Harris reference as used for the experimental Routhians
 is subtracted from the calculated ones.  

\begin{figure}[t]
\mbox{\psfig{file=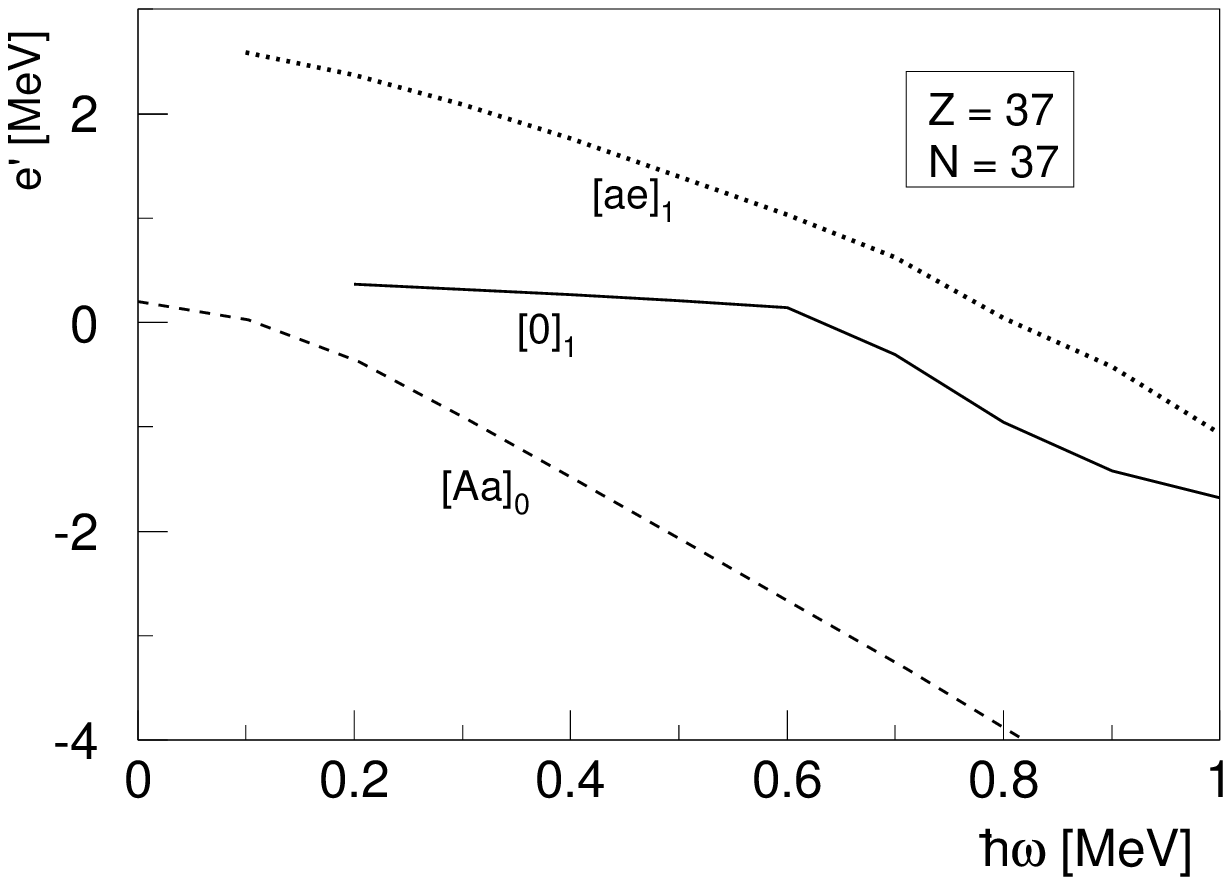,width=14cm}}
\caption{\label{rb74trs.fig}
Total routhians for  $^{74}_{37}$Rb$_{37}$ calculated by
means of the deformation optimized Woods Saxon Strutinsky method.  
 The text explains how the energy of the $T=1$ bands  relative
to the  energy of the $T=0$ ground state is fixed.
The parity and signature assignments $(\pi,\alpha)$ are:
Full lines (+,0), dashed (+,1) and
dotted (-,1). A Harris reference is subtracted.
   }
\end{figure}

It seen by comparing fig. \ref{rb74trs.fig} with fig. \ref{rb74.fig}
that the calculated spectrum now agrees much better with the data at 
high $\omega$. As compared to the CSM
calculation based on the Nilsson potential, the main differences are:
i) different (and probably better) positions of the spherical single
particle levels, ii) different deformations of the $T=0$ ($\beta\approx0.31$)
and $T=1$ bands ($\beta\approx0.39$) and iii) inclusion of the 
quadrupole pair-fields. The expression (\ref{esymm}) places $[0]_1$ somewhat
too high relative to the $T=0$ bands. The reason is that the TRS calculation
gives an excitation energy of $2.5 MeV$ for the configuration $[Aa]$ relative to
$[0]$. The fits (\ref{twodel}) 
which rather well reproduce the experimental differnce between
the $T=1$ and 0 bands in odd-odd nuclei (cf. section \ref{kr72.sec} and ref.
\cite{vogel}), give the larger value of $2\Delta=3.0 MeV$
 for $N=Z$.

\section{The role of $t=1$ proton-neutron pair-correlations}\label{pncorr.sec}

Although the construction of the intrinsic configurations within the CSM
frame looks as if there was no pn-pairing, it is implicitly included.
It is the spontaneous breaking of the isospin symmetry that permits to 
choose the orientation of the $t=1$ pair-field such that the pn-component
of the field disappears. However, in order to restore the isospin symmetry
of the total wave function the pn-pairing is absolutely necessary. 
Its strength is completely determined by the isospin symmetry.
Hence, the $t=1$ pn-pairing manifests itself by the isospin symmetry 
of the states. In this section we are going to elucidate this important
point further. 

One important consequence of the isospin symmetry is that
in $N=Z$ nuclei, the  $T=0$ configurations
like $[AB]_0$ or $[Ab]_0$ appear only once. The additional
 configurations 
obtained by exchanging quasiprotons with quasineutrons, which would 
have the same energy
if there was no pn-pairing, do not appear. More accurately one must say
that only the $T_y=0$ combinations $([AB]+[ab])/\sqrt{2}$ and
$([Ab]-[aB])/\sqrt{2}$ appear as low-lying configurations,
whereas  the $T_y\not=0$
combinations $([AB]-[ab])/\sqrt{2}$ and
$([Ab]+[aB])/\sqrt{2}$ are pushed to large energy by the pn-pair-correlations.
This symmetry restriction can be considered as a special case of a  
more general feature of  the pn-correlations.
It is a consequence of the fact that the total wavefunction has for
$T=0$ a constant probability for all orientations of $\vec \Delta$.
For small $T$, the total wave function $D^T_{TT}$ still corresponds to
a substantial probability for an orientation of $\vec \Delta$ different
from the y-direction, i.e, to a substantial pn-pair field. As a consequence, one of the two linear combinations
is  energetically favored over the other. Only for large
$T$, when the wavefunction becomes concentrated near small angles $\theta$,
the proton and neutron excitation become independent.     
This has been demonstrated in ref. \cite{fsr94} by a
systematic SM study states with the character 
 $([AB]\pm[ab])/\sqrt{2}$, which represent the two s-bands in the $h_{11/2}$
shell. It is demonstrated that for $N \approx Z$ one of the bands is pushed 
up relative to the other by the pn-interaction and the wave function
is a linear combination of the proton and neutron excitations.  
On the other hand, for  
 $Z \ll N$  the pn-interaction does not change the relative position of the 
two bands very much, which are almost pure neutron or proton excitations.
As discussed in \cite{fsr94}, the data on the alignment of $h_{11/2}$ particles
in the mass 120-130 region seems to support this prediction of the theory.
Clearly,  more detailed measurements of the rotational
spectra in nuclei near $N\approx Z$ are necessary to test this signature
of the $t=1$ pn-pair correlations.

Another consequence of the pn-interaction is the systematic enlargement
of the rotational frequency where first alignment (ab and/or AB) appears
when the intruder shell becomes more symmetrically filled. It has been
demonstrated  in ref. \cite{fs} that this effect is generated by the 
$T=1$ components of the pn-interaction.
Fig.\ref{jx.fig} illustrates this point showing a SM
calculation where we took off all the t=0 components of the 
$\delta$-interaction.
The crossing shows up at almost the same frequency as  in the calculation
with the full interaction. Hence the possible $t=0$ correlations cannot influence
the crossing in an important way. The CSM calculation 
predicts the crossing earlier than the SM.  
 Generally, the comparison 
 with the exact SM calculations in section
\ref{csmsm.sec} shows that the CSM approximation systematically underestimates
 the frequency of the alignment processes.
Since the  protons and neutrons
are assumed to independently move in a fixed rotating field,
the CSM can also not reproduce the delay of the 
the (ab) crossing caused by the presence of an A quasineutron,  
which is found  in
the full SM calculations and in the experiment as well \cite{fs}, 
The mechanism of this
effect, which seems to be of the same origin as the 
late alignment in $N\approx Z$ nuclei, is still an open question.
It is possible    that it is caused by the part of $t=1$ pn-pair
correlations that are not taken into account by the CSM approximation.

In order to check this conjecture and to disentangle it from other 
mechanisms (shape changes, for example) causing similar modifications
of the alignment processes, a 
 careful treatment of the isospin symmetry and the angular momentum
dependence of the shape and pairing degrees of freedom is necessary.
The example in section \ref{trs.sec} 
demonstrates that combining the concept of spontaneous 
breaking of isospin symmetry with careful mean-field calculations 
by means of codes that take only the pp- and nn- pairing into account
is expected to provide a good description of $N=Z$ nuclei with a strong
$t=1$ pair field. Such calculations may serve as a bench mark to 
look for $t=0$ pair correlations. 

The present paper also sheds light on the results of the recent
analysis of high spin
data in nuclei with $T_z=1/2$ and 1 \cite{kr74,rb75} by means of 
the conventional mean-field approach that does not explicitly take into
account the proton-neutron pairing. It is
stated there 
that ``the agreement between theory and experiment can be considered 
as very good...'' These results are consistent with
 our suggestion that in 
nuclei with $70<A<80$ there is strong $t=1$ pairing and
we want to point out once more our most important message:
 The fact that the mean-field theory without an explicit 
pn-pair field works well does by no means imply that there is no
 $t=1$  pn-field. On the contrary, it must have a strength comparable
with the pp- and nn-fields in order to restore the isospin symmetry. 
Hence the other statement of the paper \cite{kr74},
``we do not find clear evidence for collective pn-pairing...'' must be
taken literally. Many features of the rotational 
spectra are insensitive to the presence of the pn-field, which
 manifests itself in a rather subtle way
via isospin conservation. The discussed absence of pairs of two quasiparticle
excitations (two quasiproton and two quasineutron), which are expected
if there was no pn-pair field, is one consequence of the pn-pairing. 
The available data on 
$N=Z$ nuclei  do not  permit stringent tests of these CSM predictions so far.

In refs. \cite{kvasil,zhang} the consequences of the pn-pair correlations
are studied by introducing a pn-pair fields, the strength of which is
varied. Concerning the $t=1$ pair-fields,  
a free variation is inconsistent with the isospin conservation
which fixes the ratios between the pn-, pp- and nn-fields.

\section{Conclusions}

The cranked shell model mean-field approach has been extended to
include $t=1$ proton-neutron pairing in order to describe nuclei near
$N=Z$. The central  concept encountered is the spontaneous breaking of 
the isospin
invariance by the isovector pair-field. All orientations of the 
pair-field represent one and the same intrinsic state. One particular
choice is the direction with no pn-pair field. On the mean-field
level this permits to treat
the $N\approx Z$ nuclei essentially as if there was no pn-pairing.
Of course, the pn-field exists and it is strong. However, it is completely
determined by the isospin symmetry. It comes into play when interpreting the 
mean-field solutions as intrinsic states of the total wavefunction,
with good isospin.  As in the analogous case of spatial rotation and
deformed mean-field solutions, the isospin symmetry may be restored
with different accuracy. In this paper we have  focused on the simplest
possibility, the  limit of strong symmetry breaking. In this limit it is
well known from spatial rotations how to connect the symmetry breaking
mean-field solutions with the quantal states of good $I$ or $T$, respectively.
The resulting scheme is very simple. The spectrum is generated 
from quasiparticle excitations with no pn-pairing in the standard way.
Thus, an odd number of quasineutrons means 
odd $N$ and the same holds for protons.
The following additional rules have to be applied for $N\approx Z$:

1) The isospin is fixed by ``cranking in isospace'', i.e. 
$\langle T_z\rangle =T$.  For $T=T_z$ states, this amounts to the
 ordinary constraints in 
particle number that fix the chemical potentials $\lambda_p$ and $\lambda_n$.
 The energy of states $T_z < T$ is taken to be the
same as the isobaric analog states $T=T_z$.

2) The lowest quasiparticle excitations for the given values of 
$\lambda_p$ and $\lambda_n$ are the lowest states
 of the corresponding value of $T=\langle T_z \rangle$.

3) A term  $(\lambda_n-\lambda_p)/2$ is added to the energies.
It is a consequence of isospin conservation and, hence, a manifestation
of the pn-pairing.       

4) For the $T=0$ states in $N=Z$ nuclei,
only  quasiparticle configurations with $T_y=0$ are permitted. This 
additional symmetry restriction
excludes certain configurations that are permitted in nuclei with large $T$.

We have exactly solved the  SM problem for the model system of 
protons and neutron in an $f_{7/2}$ shell interacting via a 
$\delta$-force and exposed to an rotating 
external quadrupole potential.
The spectra for the half filled shell, $Z=3,4$ and $N=3,4$ have been
calculated as a function of the rotational frequency $\omega$. 
The rotational bands
generated in this way are compared to the mean-field theory. The mean-field 
calculations are carried out in the spirit of the 
cranked shell model approximation. The mean-field is determined for the 
$(N=Z=4)$ system at zero-frequency by solving the Hartree-Fock-Bogolubov
equations. The solution turns out to be
an isovector pair-field with a dominating monopole component but substantial
contributions from the higher multipoles. Keeping this mean-field constant,
states with finite angular momentum and isospin are generated by changing
$\omega, ~\lambda_n$ and $\lambda_p$. The structure of the exact
SM excitation spectra is reproduced by this CSM procedure, 
where the modifications 1 - 4) turn out
to be important. A fair quantitative agreement is also obtained. In particular,
the relative position of $T=0$ and $T=1$ bands is well reproduced.
The most conspicuous discrepancy is that the CSM predicts the rotational
alignment of quasiparticle pairs at a too low-frequency and is not
able to describe the delay of the neutron alignment by the presence of
odd protons in the same shell (and vice versa). 

With the aforementioned restrictions, the suggested CSM appears to be quite
 capable to describe the yrast-spectra of rotating nuclei near $N=Z$.
One expects the CSM to be  better for  realistic nuclei
than for our $f_{7/2}$ model case, because the larger number of nucleons
favors the mean-field approximation. Realistic  Hartree-Fock-Bogolubov
calculations \cite{sandhu} and SM calculations
\cite{engel,dean} point to the existence of 
 a  $t=1$ pair-field at low-spin in the mass 70 region. 
Hence we have applied our CSM
to $^{72}_{36}$Kr$_{36}$, $^{73}_{36}$Kr$_{37}$ and $^{74}_{37}$Rb$_{37}$.
Satisfactory agreement with the data is found and 
the assumption of a $t=1$ pair-field in $N\approx Z$ 
nuclei in the mass 70 -80 region is consistent with the
 existing measurements.        
In particular, the recently observed crossing between the 
$T=1$ even-spin ground band and a $T=1$ odd-spin band in  $^{74}$Rb
can be well explained within the scenario of a $t=1$ pair-field.
The close energy of the two bandheads, which is the reason for the crossing,
is a consequence of the near cancelation of two larger energies,
twice the pair-gap, by which the the $T=0$ band is pushed up, and the
``iso-rotational'' energy (symmetry + Wigner energy), by which the $T=1$ band
is pushed up. This compensation is a quite general feature of the 
odd-odd $N=Z$ nuclei, as pointed out in ref. \cite{vogel}.
 
Finally, it should be pointed out that isospin is not exactly conserved.
The  Coulomb potential breaks this symmetry. This has the interesting 
consequence that the isovector pair-field does not completely rotate freely
in isospace (i.e. the wavefunction is not exactly 
a $D$ function), but it feels
the Coulomb field like an external potential, which prefers a certain
orientation. This effect and other
manifestations of isospin symmetry in the rotational spectra 
are very interesting subjects for future studies. In particular, 
measuring the 
differences between the moments of inertia and alignment processes of isobaric
analog bands will shed new light on the old question of isospin purity in  
heavier nuclei.


%

\newpage
\onecolumn

\newpage

\newpage

\newpage

\newpage

\newpage

\newpage

\newpage

\newpage

\newpage

\newpage

\newpage


\begin{thebibliography}{99}
\bibitem{ring} P. Ring and H. J. Mang, Phys. Rev. Lett. {\bf 33}, 1174 (1974)

\bibitem{bf79} R. Bengtsson and S. Frauendorf, 
Nucl. Phys. {\bf A327}, 139 (1979)

\bibitem{devoigt} M. J. A. de Voigt, J . Dudek, Z. Szymanski,
 Rev. Mod. Phys. {\bf 55}, 949 (1983)

\bibitem{bg} R. Bengtsson and J.D. Garrett, Int. Rev. Nucl. Phys.
{\bf 2}, 193 (1984)

\bibitem{pairing} Y. R. Shimizu, J. D. Garrett, R. A. Broglia, M. Gallardo,
E. Vigezzi, Rev. Mod. Phys. {\bf 61}, 131 (1989)

\bibitem{len96}  S.M. Lenzi et al., Z. Phys. {\bf A354}, 117 (1996)

\bibitem{lea97}  C.D. O'Leary, M.A. Bentely, D.E. Appelbe, D.M. Cullen,
S. Erturk, R. Bark and A. Maj, Phys. Rev. Lett. {\bf 79}, 4349 (1997)

\bibitem{kr72} G. de Angelis et al., Phys. Lett. {\bf B415}, 217 (1997)


\bibitem{srn90} J.A. Sheikh, M.A. Nagarajan and N. Rowley,
Phys.  Lett. {\bf B240}, 11 (1990)

\bibitem{she90} J.A. Sheikh, N. Rowley, M.A. Nagarajan and H.G. Price,
Phys. Rev. Lett. {\bf 64}, 376(1990)

\bibitem{fsr94}  S. Frauendorf, J.A. Sheikh and N. Rowley,
Phys. Rev. {\bf C50}, 196 (1994)

\bibitem{mueller} E. M. M\"uller, K. M\"uhlhans, K. Neergard, and 
U. Mosel, Nucl. Phys. {\bf A383} (1982) 233 

\bibitem{pnr89} K.F. Pal, M.A. Nagarajan and N. Rowley,
 Nucl. Phys. {\bf A500}, 221 (1989)

\bibitem{goodman} A. L. Goodman, Adv. of Nucl. Phys. {\bf 11}
(1979) 263

\bibitem{sandhu} T. S. Sandhu, and M,. L. Rustgi, 
Phys. Rev. {\bf C 14}, 675 (1976)

\bibitem{satula} W. Satula and R. Wyss, Phys. Lett {\bf B 393}, 1 (1997)

\bibitem{engel} J. Engel et al., Phys. Lett. {\bf B 389}, 211 
(1996)

\bibitem{dean} D. J. Dean, S. E. Koonin,
 K. Langanke and P. B. Radha, Phys. Lett. {\bf B 399}, 1 (1997) 

\bibitem{ringschuck}  P. Ring, P Schuck, The Nuclear Many Body Problem,
(Springer N, 1980), p. 244 ff

\bibitem{cg} H. T. Chen and A. Goshwami, Phys. Lett. {\bf 24B}, 257 (1967)

\bibitem{T0} H. H. Wolter, A. Faessler, and P. U. Sauer, Nucl. Phys.
{\bf A167}, 108 (1971) 

\bibitem{goodgosw}  A. L. Goodman et al.,
Phys. Lett. {\bf B26} (1968) 260
 
\bibitem{GO} A. L. Goodman, Nucl. Phys. {\bf A186}, 475 (1972)

\bibitem{sw} J.A. Sheikh and R. Wyss (to be published)

\bibitem{fst0}   S. Frauendorf and J.A. Sheikh (to be published)

\bibitem{bm2} A. Bohr and B. Mottelson, Nuclear Structure II, (W. A. Benjamin,
1975)

\bibitem{camiz} P. Camiz, A. Covello and M. Jean,
 Nouvo Cim. {\bf 42 B}, 199 (1966)

\bibitem{ginocchio} J. N. Ginocchio and J. Weneser, Phys. Rev. {\bf 170}, 859
(1968)

\bibitem{rb74}  D. Rudolph et al., Phys. Rev. Lett. 76 (1996) 376

\bibitem{vogel} P. Vogel, preprint nucl-th/9805015 and earlier references
cited


\bibitem{tac} S. Frauendorf, Nucl. Phys. {\bf A557}, 259c (1993)

\bibitem{nilsson} S. G. Nilsson and I. Ragnarsson, Shapes and Shells in
Nuclear Structure, Cambridge University Press, (1995)

\bibitem{galeriu} D. Galeriu, D. Bucurescu and M. Ivascu, J. Phys. 
{\bf G12}, 329 (1986)

\bibitem{kr74} D. Rudolph et al., Phys. Rev. {\bf C56}, 98 (1997)

\bibitem{rb75} C. J. Gross et al., Phys. Rev. {\bf C56}, R591  (1997)

\bibitem{meng} S. Frauendorf and J. Meng Z. Phys. {\bf A356},  263 (1996)

\bibitem{kr73} S. Freund et al., Phys. Lett. {\bf B 302}, 167 (1993)

\bibitem{br72} Nucl. Data Sheets

\bibitem{fs}   S. Frauendorf and J.A. Sheikh, preprint nucl-th/9806094 

\bibitem{rb74trs} W. Satula, private communication

\bibitem{kvasil} J. Kvasil, A. K. Jain, and R. Sheline, Czech. J. Phys.
{\bf 40}, 278 (1990)

\bibitem{zhang} K. Kaneko and Jing-je Zhang, Phys. Rev. {\bf C 57}, 1732 (1998)

\end{thebibliography}
\end{document}